\documentclass[12pt,manuscript,authoryear]{aastex}

\usepackage{graphicx}
\usepackage{natbib}
\usepackage{epstopdf}

\usepackage{color}
\usepackage{ulem}
\definecolor{darkred}{rgb}{0.8,0.1,0.1}

\newcommand{\rvec}{\mbox{\boldmath$r$}}

\newcommand{\vvg}{\mbox{\boldmath$V_{\rm g}$}}

\newcommand{\vvc}{\mbox{$\mbox{\boldmath$V$}_{\rm c}$}} 

\newcommand{\gtsimeq}{\raisebox{-0.6ex}{$\, \stackrel{\raisebox{-.2ex}%
{$\textstyle >$}}{\sim}\,$}}
\newcommand{\ltsimeq}{\raisebox{-0.6ex}{$\, \stackrel{\raisebox{-.2ex}%
{$\textstyle <$}}{\sim}\,$}}

\newcommand{\app}{\ensuremath{\sim} }

\begin{document}

\title{Chondrule Formation in Bow Shocks around Eccentric Planetary Embryos}
  
\author{{\it Short Title: Planetary Embryo Bow Shocks ~~~~Article Type: Journal}}

\author{Melissa~A.~Morris}
\affil{School of Earth and Space Exploration, Arizona State University,
        P.~O.~Box 871404, Tempe, AZ 85287-1404}
\email{melissa.a.morris@asu.edu}

\and

\author{Aaron C. Boley}
\affil{Department of Astronomy University of Florida, Gainesville, FL 32611}

\and

\author{Steven J. Desch}
\affil{School of Earth and Space Exploration, Arizona State University,
        P.~O.~Box 871404, Tempe, AZ 85287-1404}

\and

\author{Themis Athanassiadou}
\affil{School of Earth and Space Exploration, Arizona State University,
        P.~O.~Box 871404, Tempe, AZ 85287-1404}

\begin{abstract} 

Recent isotopic studies of Martian meteorites by Dauphas \& Pourmond (2011) have
established that large ($\sim 3000$ km radius) planetary embryos existed in the solar 
nebula at the same time that chondrules - millimeter-sized igneous inclusions found in 
meteorites - were forming.
We model the formation of chondrules by passage through bow shocks around such
a planetary embryo on an eccentric orbit.  
We numerically model the hydrodynamics of the flow, and find that such large bodies 
retain an atmosphere, with Kelvin-Helmholtz instabilities allowing mixing of this 
atmosphere with the gas and particles flowing past the embryo.
We calculate the trajectories of chondrules flowing past the body, and find that they
are not accreted by the protoplanet, but may instead flow through volatiles outgassed
from the planet's magma ocean. 
In contrast, chondrules {\it are} accreted onto smaller planetesimals. 
We calculate the thermal histories of chondrules passing through the bow shock.
We find that peak temperatures and cooling rates are consistent with the formation of the
dominant, porphyritic texture of most chondrules,
assuming 
a modest enhancement above the likely solar nebula average value of chondrule 
densities (by a factor of 10), attributable to settling of chondrule precursors to the
midplane of the disk or turbulent concentration.
We calculate the rate at which a planetary embryo's eccentricity is damped and 
conclude that a single planetary embryo scattered into an eccentric orbit can, over
$\sim 10^5$ years, produce $\sim 10^{24} \, {\rm g}$ of chondrules.
In principle, a small number ($1 - 10$) of eccentric planetary embryos can melt the
observed mass of chondrules in a manner consistent with all known constraints. 

\end{abstract}

\keywords{meteorites, meteors, meteoroids; protoplanetary disks; radiative transfer; 
shock waves; instabilities; planet-disk interactions}

\label{firstpage}

\section{Introduction}

Chondrules are submillimeter to millimeter-sized, mostly Fe-Mg silicate 
spherules found within chondritic meteorites.  
Chondrule precursors were heated to high temperatures and melted, 
and then slowly cooled and crystallized to form chondrules,
while they were independent, free-floating objects in the early solar nebula.
A wealth of elemental and isotopic compositional measurements, as well as 
mineralogical and petrological information, potentially place important constraints 
on the melting and cooling of chondrules.
The physical conditions and processes acting in the early solar nebula could
be deciphered, if the chondrule formation mechanism could be understood.

Various chondrule formation models have been proposed.
These are assessed against their ability to match the following meteoritic constraints: 
(1) Models must explain why most chondrules appear to have formed only millions of years 
after the oldest solids formed in the solar nebula (calcium-rich, aluminum-rich inclusions, otherwise known as CAIs). 
Radiometric (Al-Mg and Pb-Pb) dating shows that nearly all extant chondrules formed 
$\approx 1.5 -4 \, {\rm Myr}$ after CAIs, with the majority forming 2-3 Myr after CAIs 
(Kurahashi et al. 2008; Villeneuve et al. 2009).  
(2) Models must explain the high number density of chondrules in the formation region. 
About 5\% of chondrules are compound chondrules, chondrules that fused together while
still partially molten (Gooding \& Keil 1981; Ciesla et al.\ 2004a).  
Based on this frequency and an assumed relative velocity $100 \, {\rm cm} \, {\rm s}^{-1}$, 
a number density of chondrules $\sim 1 - 10 \, {\rm m}^{-3}$ has been inferred 
(Gooding \& Keil 1981; Ciesla et al.\ 2004a). 
Cuzzi \& Alexander (2006) also require a density $> 10 \, {\rm m}^{-3}$ so that the 
partial pressure of volatiles following chondrule melting remains high enough to suppress 
substantial loss of further volatiles. 
This density is significantly higher than the average density of chondrules in
the solar nebula.
Based on a silicate-to-gas mass fraction $0.5\%$ (Lodders 2003) and the observation
that chondrules comprise 75\% of the mass of an ordinary chondrite (e.g., Grossman 1988),
we expect the chondrules-to-gas ratio to be $0.375\%$ (although this value may be lower if a large percentage of the available solids are locked up in planetary embryos).
Assuming a chondrule mass $m_{\rm c} = 3.7 \times 10^{-4} \, {\rm g}$, consistent with a chondrule of radius $300 \, \mu{\rm m}$ and internal density
$3.3 \, {\rm g} \, {\rm cm}^{-3}$, and a nebula gas density 
$1 \times 10^{-9} \, {\rm g} \, {\rm cm}^{-3}$ (Desch 2007), we would expect a chondrule
number density $0.01 \, {\cal C} \, {\rm m}^{-3}$, where the chondrules-to-gas mass
ratio is defined to be $3.75 \times 10^{-3} \, {\cal C}$, with ${\cal C}$ representing an additional concentration factor.
Here, the chondrule precursor radius of $300\mu$m is chosen to be consistent with chondrules found in most chondrites.   It is reasonable, however, to assume that there was a size distribution between matrix-sized grains and large chondrule precursors in the solar nebula prior to a chondrule-forming event.  Morris \& Desch (2012) find that cooling rates are  increased slightly for simple size distributions, but not enough to affect the general results of the shock model.  We therefore consider only a single size for simplicity.
In the chondrule-forming region, chondrules appear to have been concentrated by large
factors, up to 100 times
their background density (for a relative velocity of $100 \, {\rm cm} \, {\rm s}^{-1}$). 
(3) Models must allow for simultaneous chondrule formation over large regions 
($> 10^3 \, {\rm km}$ in size) 
to limit diffusion of volatiles away from chondrules,
in order to keep the partial pressures of the volatiles high
enough to suppress isotopic fractionation of these volatiles (Cuzzi \& Alexander 2006). 
Likewise, the existence of compound chondrules argues for regions larger than the mean
free path of a chondrule through a cloud of chondrules, which  is 100 - 1000 km for the densities quoted 
above.
(4) Models must explain the relatively slow cooling of chondrules over hours. 
Comparison of chondrule petrologic textures with experimental analogs show 
that the 80\% of chondrules in ordinary chondrites that have porphyritic 
textures (Gooding \& Keil 1981) cooled from their peak temperatures at rates 
$\sim 10 - 10^{3} \, {\rm K} \, {\rm hr}^{-1}$ and crystallized in a matter
of hours (Desch \& Connolly 2002; Hewins et al.\ 2005; Connolly et al. 2006; 
Lauretta et al. 2006). 
(5) The partial pressure of Na may have been considerably higher than is
easily explained by volatile loss from chondrules. 
Alexander et al.\ (2008) 
recently reported high mass fractions
$\approx 0.010 - 0.015 \; {\rm wt}\%$ of Na$_2$O in the cores of olivine phenocrysts 
of Semarkona chondrules, indicating that the olivine melt contained substantial Na.  
The implied partial pressures of Na vapor in the chondrule-forming region are 
$\sim 10^{-5} \, {\rm bar}$.  If this vapor is supplied by the chondrules
themselves, chondrule concentrations ${\cal C} > 10^5$ would be required. 
Such a high chondrule concentration would have significant and even incredible 
physical consequences, but it has been difficult to explain the high Na concentrations  
in the chondrule olivine by any other method (Alexander et al.\ 2008).
Many additional constraints on chondrule formation exist as well 
(see Desch et al.\ 2012, accepted). 
The model that appears most consistent with the constraints on chondrule 
formation is the nebular shock model, in which chondrules are melted by
passage through shock waves in the solar nebula gas (Desch et al.\ 2005, 
2010, 2012).
Of the sources of shocks considered to date, two were found to be most 
plausible by Desch et al.\ (2005): large-scale shocks driven by gravitational 
instabilities (e.g., Boss \& Durisen 2005; Boley \& Durisen 2008), 
and bow shocks driven by planetesimals on eccentric orbits 
(e.g., Hood 1998; Ciesla et al.\ 2004b; Hood et al.\ 2009; Hood \& Weidenschilling 
2011).

A planetesimal or other body on an eccentric orbit will inevitably drive a bow 
shock in front of it, because its orbital velocity necessarily differs from that 
of the gas, which remains on circular orbits with nearly the Keplerian velocity $v_{\rm K}$.
A body with eccentricity $e$ will have a purely azimuthal velocity at aphelion 
or perihelion, but one that differs from the local Keplerian velocity by an amount 
$\sim e v_{\rm K} / 2$.  
At other locations in its orbit the protoplanet also has a radial velocity.
If the body's orbit is inclined, this also enhances the relative velocity.
The combined effect is that the body's velocity relative to the gas can remain a 
substantial fraction of the Keplerian velocity 
($v_{\rm K} \sim 20 \, {\rm km} \, {\rm s}^{-1}$ at 2 AU) over its entire orbit. 
As one plausible example, which we discuss further in section 6.1, we show in Figure 1 the velocity difference 
$V_{\rm s}$ between the gas and solid body on an eccentric orbit with semi-major 
axis $a = 1.25 \, {\rm AU}$, eccentricity $e = 0.2$ and inclination $i = 15^{\circ}$.
The relative velocity is seen to vary over the orbit but it remains 
$> 3 \, {\rm km} \, {\rm s}^{-1}$ at all times. 
As the typical sound speed in the gas is $< 1 \, {\rm km} \, {\rm s}^{-1}$, a bow shock is 
the inevitable result of an eccentric planetary body. 
The authors above (Hood 1998; Ciesla et al.\ 2004b; Hood et al.\ 2009; Hood
\& Weidenschilling 2011) have argued that planetesimals, bodies up to several $\times 10^2 \, {\rm km}$
in radius, on eccentric orbits, would be common enough to drive sufficient shocks to 
melt the observed mass of chondrules, which is $2 \times 10^{-4} \, M_{\oplus}$ 
(Grossman 1988). 

Despite their presumed ubiquity, planetesimal bow shocks have not been favored as the 
chondrule-forming mechanism.
Because chondrules are associated with asteroids and formation in the asteroid
belt, only planetesimals $\ltsimeq 10^{3} \, {\rm km}$ in diameter have been
considered so far.
For such bodies, the size of the heated region, which is necessarily comparable to
the size of the planetesimal itself, is probably too small to satisfy the
above constraints on the chondrule-forming region.
More importantly, the small size of the heated region means chondrules are never far 
from cold, unshocked gas to which they can radiate and cool.
The optically thin geometry is expected to lead to very fast cooling rates, and 
matching the cooling rates necessary to reproduce the dominant porphyritic texture in particular,
$\ltsimeq 10^3 \, {\rm K} \, {\rm hr}^{-1}$, requires physical conditions that appear 
extreme (e.g., Ciesla et al.\ 2004b; Morris et al.\ 2010a,b). 
Intuitively, suppression of the radiation that cools the chondrules requires requires 
the optical depth $\tau$ between the chondrule-forming region and the unshocked gas to 
be $\gtsimeq 1$.
In their treatment of the problem, Morris et al.\ (2010a,b) found
that optical depths $\tau \gtsimeq 0.1$ led to cooling rates
$\sim 10^3 \, {\rm K} \, {\rm hr}^{-1}$, just barely consistent with
the constraints.  Our treatment of chondrule cooling here (see \S 4)
suggests slightly slower cooling, but the cooling rates
$\sim 100 \, {\rm K} \, {\rm hr}^{-1}$ consistent with porphyritic
textures require optical depths $\tau \sim 0.1$.  The optical depth is a function of the opacity and the size of the region.
Dust grains that might otherwise increase the opacity are found to vaporize in shocks 
strong enough to melt chondrules (Morris \& Desch 2010).
The opacity of chondrules themselves, at their background levels, does not yield 
$\tau > 0.1$ unless the size of the region is $l \gg 3 \times 10^{4} \, {\rm km}$.
Even if chondrules are concentrated by the considerable factor of 100 times their 
background level, so that $n_{\rm c} = 1 \, {\rm m}^{-3}$, planetesimals at least several
hundred kilometers in radius are required to yield marginally consistent cooling rates
(Ciesla et al.\ 2004b; Morris et al.\ 2010a,b).
To the extent that mass is locked up in large planetary bodies rather than dust and
chondrule-sized objects, even higher concentrations of chondrules above background
levels are required. 

Another difficulty of the planetesimal bow shock model, one that has not been discussed
previously, is that due to the small size of the planetesimal, chondrule precursors 
passing close enough to the body to be melted as chondrules are likely to be accreted by the 
planetesimal.
Unlike the gas, whose velocity is immediately altered at the shock front to flow around
the body, chondrules have sufficient inertia that they retain their pre-shock velocity 
until they have collided with their own mass of gas. 
Chondrules can be deflected around the body driving the bow shock only by moving 
a minimum length through the gas past the shock front.  For typical parameters this
length is $\approx 300 \, {\rm km}$. 
The stand-off distance between the bow shock and a 500 km-radius planetesimal driving it is typically 
$\ltsimeq 100 \, {\rm km}$ (see \S 2), meaning that accretion onto the planetesimal is 
probable.
This is problematic because chondrules are expected to mix with cold nebular components 
before accretion, and a given chondrite will contain chondrules with a wide variety of ages. 

Chondrules melted in bow shocks around bodies {\it larger} than planetesimals may conform 
better to the constraints on chondrule cooling rates and the size of the chondrule forming region. 
The terrestrial planets have long been recognized to have formed from planetary
embryos, bodies $> 1000 \, {\rm km}$ in diameter or even much larger, and it has
been recognized that these embryos could have formed in a few Myr or less 
(Wetherill \& Stewart 1993; Weidenschilling 1997). 
Mars itself is considered a starved planetary embryo (Chambers \& Wetherill 1998),
which allows the timing of embryo formation to be fixed. 
Previous Hf-W isotopic analyses had suggested Mars had differentiated within
10 Myr of CAI formation (Nimmo \& Kleine 2007).
More recently, the analysis of Hf and W, in conjunction with Th, by Dauphas \& 
Pourmond (2011) suggests that Mars accreted 50\% of its mass (and 80\% of its
radius) within 2 Myr of CAIs. 
Formation of the terrestrial planets requires dozens of such embryos.  
The timing of Mars's formation establishes that the solar nebula contained not just 
planetesimals, but {\it protoplanets}, during the epoch of chondrule formation.

Moreover, some of these planetary embryos quite plausibly found themselves on eccentric 
orbits at least some of the time.
Mutual scattering events are common in N-body simulations (Chambers 2001; Raymond et. al 2006). 
One recent model for the origin of Mars in particular suggests that it suffered close 
encounters with other bodies in an annulus interior to 1 AU, and was scattered into an inclined, 
eccentric orbit with aphelion at 1.5 AU (Hansen 2009).  
Alternatively, Walsh et al.\ (2011) have simulated the early migration of Jupiter and
shown it leads to a truncated inner disk (consistent with the Hansen 2009 model) and an
excess of dynamically excited planetary embryos (at least 5 in number) at 1.0 AU, from
which they propose Mars formed. 
The true extent to which embryos were scattered is not well known, but is a common feature of 
planet formation models.

The recent confirmation by Dauphas \& Pourmand (2011) that large planetary embryos did exist 
at the time of chondrule formation, the expectation that such planetary embryos may be scattered
onto eccentric orbits, and the larger size of the chondrule forming region around planetary 
embryos strongly motivate us to consider chondrule formation in bow shocks around planetary 
embryos several 1000 km in radius.
Despite the obvious similarities to chondrule formation in planetesimal bow shocks, chondrule
formation in planetary embryo bow shocks differs in several significant ways. 
First, as the planetary radius and size of the chondrule forming region increase above a 
threshold of about 1000 km (for the embryo radius), the region of heated chondrules becomes optically thick, fundamentally
altering the chondrule cooling rates. 
Second, chondrules flowing around bodies larger than about 1000 km in radius can dynamically recouple with 
the gas and avoid hitting the embryo's surface.
Chondrules in planetesimal bow shocks, if their impact parameters are less than the planetesimal
radius, are likely to be accreted by the planetesimal. 
A third fundamental difference is that planetary embryos larger than about 1000 km in radius
are able to accrete a primary atmosphere from the nebula and retain secondary outgassed atmospheres. 
The escape velocity from a body with radius $R_{\rm p}$ and Mars-like density ($\rho$ = 3.94~g~cm$^{-3}$) is 
$1.5 \, (R_{\rm p} / 1000 \, {\rm km})^{1/2} \, {\rm km} \, {\rm s}^{-1}$, compared to
a typical sound speed $0.7 \, {\rm km} \, {\rm s}^{-1}$ (at 150 K).
The atmosphere retained by the larger planetary embryo will affect the trajectories
of gas and chondrules flowing past it, and will also alter the chemistry of the chondrule
formation environment. 
We are motivated to study chondrule formation in planetary embryo bow shocks because of
these fundamental differences. 

Here we present new calculations of the melting of chondrules in bow shocks around large 
protoplanets on eccentric orbits. 
We show that for reasonable parameters, chondrule formation in planetary embryo bow shocks is
consistent with the major constraints listed above. 
We show that for the largest eccentricities and inclinations we consider plausible
the shocks speeds of $6 - 8 \, {\rm km} \, {\rm s}^{-1}$ needed for chondrule formation 
(Morris et al.\ 2010a,b; \S 4) are sustained for much of the protoplanet's orbit.
For the first time, we follow the trajectories of chondrules formed in bow shocks and 
demonstrate that they are not accreted by the protoplanet, whereas chondrules formed in 
bow shocks around planetesimals {\it are} accreted. 
Based on the spatial distribution of chondrules, we calculate the optical depths and 
calculate chondrule cooling rates. 
We show that the large scale of planetary embryo bow shocks yields chondrule cooling 
rates consistent with meteoritic constraints. 
We show that chondrules formed in planetary embryo bow shocks do pass through what is 
essentially the planet's upper atmosphere, and therefore may be exposed to high vapor pressures 
of species outgassed from the protoplanet's magma ocean, including ${\rm H}_{2}{\rm O}$ and Na
vapor.
We compute the rate at which a protoplanet's eccentricity and inclination are damped, and
show that the protoplanet's orbit remains eccentric and can drive shocks for $\sim 10^5 \, {\rm yr}$.
We find that a single scattered planetary embryo can process up to $10^{24} \, {\rm g}$ of 
chondrules.

\section{Hydrodynamics of Planetary Embryo Bow Shocks}

We have numerically investigated the hydrodynamics of a bow shock around a planetary embryo
using the adaptive-mesh, Eulerian hydrodynamics code, FLASH (Fryxell et al.\ 2000).
We consider a stationary protoplanet on a cylindrical (axisymmetric) 2-D grid.
We simulate the relative motion of the protoplanet through the gas by imposing a supersonic
inflow of gas along the symmetry axis, originating at the $-z$ boundary. 
Our assumed boundary conditions allow for inflow on the $-z$ boundary, reflection on the symmetry axis, 
and diode (zero-gradient) boundary conditions on the other boundaries.
We consider two physical cases: that of a planetesimal and one of a planetary embryo. 

In the first case of a planetesimal, we consider a body with mass $M_{\rm p} = 2 \times 10^{24} \, {\rm g}$,
a radius of $R_{\rm p} = 500 \, {\rm km}$, yielding a density of $3.80 \, {\rm g} \, {\rm cm}^{-3}$ (similar to Mars's density).  
The computational domain has radius 16,000 km and vertical extent 32,000 km.
The simulation is run with 7 levels of refinement, resulting in our best grid resolution of 32~km = $X_{\rm size}/[N_{X_{\rm block}}\cdot 2^{(\rm levels\;of\;refinement - 1)}]$, where $N_{X_{\rm block}}$ = 8.  
In the second case of a planetary embryo, 
motivated by the results of Dauphas \& Pourmand (2011), we initialize our simulations with a 
protoplanet of mass $M_{\rm p} = 3.20 \times 10^{26} \, {\rm g}$ (half of Mars's mass) and a 
radius $R_{\rm p} = 2720 \, {\rm km}$ (80\% of Mars's radius),  
with density $3.80 \, {\rm g} \, {\rm cm}^{-3}$.

Once again, the computational domain has radius 16,000 km and vertical extent 32,000 km, and the simulation is run with 8 levels of refinement.
In both cases the body is embedded in a non-radiating ${\rm H}_{2}$ / He gas with uniform 
initial density $\rho_{0} = 1.0 \times 10^{-9} \, {\rm g} \, {\rm cm}^{-3}$, pressure 
$P_{0} = 5.3 \times 10^{-6} \, {\rm atm}$, temperature $T_{0} = 150 \, {\rm K}$, and assumed 
adiabatic index $\gamma = 7/5$, appropriate for the solar nebula at around 2 AU (Desch 2007). 
The gravitational pull of the protoplanet was simulated using a point-mass gravity module.

Before we initialize the inflow, we allow the planetary embryo to acquire a primary atmosphere
of ${\rm H}_{2}$ and He.
It is straightforward to show that accretion of adiabatic gas will lead to a spherically
symmetric profile 
\begin{equation}
\rho(r) = \rho_{0} \left[1 + \frac{ r_{\rm g} }{ r } \right]^{1 / (\gamma-1)},
\end{equation}
where $r$ is the distance from the protoplanet center, 
$r_{\rm g} = (\gamma-1) (G M_{\rm p} \rho_{0} / P_{0}) / \gamma$ 
is akin to a scale height.
For the case of the planetesimal, 
$r_{\rm g} \approx 71 \, {\rm km}$, and the atmosphere
that develops is very thin and hardly resolveable by our best grid resolution 
of 32 $\, {\rm km}$.
Physically it behaves as part of the planetesimal surface in our runs. 
For the case of the planetary embryo, $r_{\rm g} \approx 11,500 \, {\rm km}$.
At the protoplanet {\it surface} the density is expected to reach 
$6.2 \times 10^{-8} \, {\rm g} \, {\rm cm}^{-3}$, the pressure to reach $1.74 \times 10^{-3} \, {\rm bar}$,
and the temperature to reach $780 \, {\rm K}$, values that are 62.5, 330.0, and 5.2 times their background values,
respectively. 
The effective scale height near the protoplanet surface is $\approx 1000 \, {\rm km}$,
and is easily resolved by our best grid resolution of 32 km.
The primary atmosphere acquired by the planetary embryo in our simulations matched this solution within a few percent. 
The mass of the atmosphere inside 1000 km of the protoplanet surface is very nearly
$3 \times 10^{18} \, {\rm g}$, also in accord with analytical estimates. 
We do not directly simulate the effects of a secondary atmosphere of species outgassed from the
protoplanet's magma ocean, which would be orders of magnitude denser, but we expect 
it would behave very similarly dynamically. 

We begin our simulation of the bow shock with the protoplanet and primary atmosphere, and introduce gas streaming up from the lower $z$ boundary with density 
$\rho_{0} = 1 \times 10^{-9} \, {\rm g} \, {\rm cm}^{-3}$ and upward velocity 
$V_{\rm s} = 8 \, {\rm km} \, {\rm s}^{-1}$.
We chose this density to represent conditions of the chondrule-forming region, presumed to 
be at 2-3 AU (but see \S 6).   However, disk models differ on gas densities in the protoplanetary disk.  For example, 
the density $\rho_{0} = 1 \times 10^{-9} \, {\rm g} \, {\rm cm}^{-3}$ is achieved in a
minimum-mass solar nebula model (Hayashi 1981) at 1.25 AU (depending on the scale
height), or at several (up to 5) AU in more massive disk models (Desch 2007). 
We chose the relative velocity of $8 \, {\rm km} \, {\rm s}^{-1}$ based on the results of Morris \& Desch (2010), who showed that this speed was necessary to achieve chondrule melting.
Such a large relative velocity may appear to be extreme, but such values are expected to be met episodically throughout planet formation, even though the typical relative velocity is expected to be much lower.  A velocity difference of $8 \, {\rm km} \, {\rm s}^{-1}$ between the body and the gas
can occur for a range of eccentricities and inclinations, which will be explored in more detail in Section 6.  Here, consider the example of a body on an orbit with a semi-major axis $a = 1.25$ AU and $e = 0.29$.  This body will produce a $\sim 8$ km/s bow shock at quadrature ($a\sim 1.1$ AU), which is a plausible site for the formation of some chondrules. Compared with the total population of embryos, such  high relative  velocities may be rare, but when they do occur, they can be very significant, as we will show. 


In both the planetesimal and planetary embryo cases, the evolution of the gas is marked by 
a reverse shock bouncing off the body surface and multiple shocks reflecting off each
other at the symmetry axis.
Within a few crossing times (i.e., a few $\times 10^3 \, {\rm s}$) for the planetesimal case, 
and a few $\times \sim 10^4 \, {\rm s}$ for the planetary embryo case), 
a quasi steady-state bow shock structure is reached.  An animation showing the case for the protoplanet is shown in Figure~\ref{movie}.
Contour plots of gas densities $4 \times 10^{4} \, {\rm s}$ after the introduction of supersonic
material on the grid is shown in Figures~\ref{fig:streamlines}a and~\ref{fig:streamlines}b.
In both panels the gas density is depicted in blue, with the darkest levels referring to densities 
of $2.5 \times 10^{-10} \, {\rm g} \, {\rm cm}^{-3}$ and the brightest levels to 
$2.5 \times 10^{-8} \, {\rm g} \, {\rm cm}^{-3}$.
At the bow shock itself the density increases by a factor of 6 at the shock front, from 
$1 \times 10^{-9} \, {\rm g} \, {\rm cm}^{-3}$ to 
$6 \times 10^{-9} \, {\rm g} \, {\rm cm}^{-3}$, as expected for a $\gamma = 7/5$ gas, in a 
strong shock.


We observed that the bow shock around the planetesimal quickly achieved steady state, but as 
is evident from Figure~\ref{fig:streamlines}b, steady-state is not formally achieved around the 
protoplanet.
This is observed by the varying density in the wake of the protoplanet, and the slightly
irregular structure of the bow shock (e.g., the kink at $x = 6000 \, {\rm km}$).
We attribute this to the boundary layer that exists between the primary atmosphere and the post-shock 
flow of gas past the protoplanet. 
Inside this boundary, evident at a distance $\approx 1000 \, {\rm km}$ from the protoplanet
surface, the densities are considerable higher $(> 10^{-8} \, {\rm cm}^{-3}$) than the 
densities $\approx 6 \times 10^{-9} \, {\rm g} \, {\rm cm}^{-3}$ between this boundary and the bow shock.
This boundary is unstable to the Kelvin-Helmholtz (KH) instability, and we observed frequent shedding
of vortices and rolls of material at this interface; these rolls are evident in Figure~\ref{fig:streamlines}b
at regular ($\approx 2000 \, {\rm km}$) intervals.
Reflections of sound waves off of these rolls is responsible for the irregular structure at the 
bow shock.

Eventually these KH rolls will strip away the atmosphere surrounding the protoplanet.
Numerical simulations of the stripping of gravitationally bound gas by KH rolls at the surfaces of 
protoplanetary disks by Ouellette et al.\ (2007) suggest that the efficiency of stripping is about 
1\%, meaning that the mass of gas that is stripped is about 1\% of the mass of the impacting gas.
Based on the inflow of gas $\rho_{0} \, \pi R_{\rm p}^2 \, V_{\rm s}$, we estimate a stripping 
rate $2 \times 10^{12} \, {\rm g} \, {\rm s}^{-1}$. 
This suggests that loss of the primary atmosphere would take $> 10^6 \, {\rm s}$, hundreds of times
longer than the duration of our simulation.  
We were able to quantify this stripping rate via a ``dye" calculation in which we kept 
track separately of material accreted before the supersonic flow of gas on the grid.
We found that after an initial loss of gas, the primary atmosphere was being lost on timescales
$\sim 10^{5} \, {\rm s}$, suggesting a stripping rate $> 10^{13} \, {\rm g} \, {\rm s}^{-1}$. 
The existence of this interface between an atmosphere and the post-bow shock flow is a fundamental
difference between the planetesimal and protoplanet cases.

\section{Chondrule Trajectories in Planetary Embryo Bow Shocks} 

Having established the dynamics of the gas and the structure of the 
bow shock surrounding the planetary embryo, we now turn our attention
to the trajectories of chondrules and their precursors in the flow. 
In general, the particles are dynamically coupled to the gas, except 
in the vicinity of the bow shock, when the gas suddenly changes velocity over a few
molecular mean free paths (i.e., over a few meters).
Particles will dynamically recouple with the gas on a timescale comparable
to the aerodynamic stopping time,
\begin{equation}
t_{\rm stop} = \frac{ \rho_{\rm s} a_{\rm c} }{ \rho_{\rm g} c_{\rm s} },
\end{equation}
where $\rho_{\rm s} \approx 3 \, {\rm g} \, {\rm cm}^{-3}$ is the particle
internal density, $a_{\rm c} \approx 300 \, \mu{\rm m}$ is its radius, 
$\rho_{\rm g}$ the post-shock 
gas density, and $c_{\rm s}$ the post-shock gas sound speed.
For $\rho_{\rm g} = 6 \times 10^{-9} \, {\rm g} \, {\rm cm}^{-3}$ and 
$T_{\rm g} \approx 2500 \, {\rm K}$, the stopping time is 
50 seconds. 
If the planetary body is traveling through the gas at a speed
$V_{\rm s} = 8 \, {\rm km} \, {\rm s}^{-1}$, particles will move past
the shock front 
several hundred km
before they dynamically couple to the gas.
Based on the numerical simulations above, this lengthscale is less than or comparable 
to the distance between the shock front and the planetary bodies themselves.
The expectation is that particles passing through a bow shock will be accreted onto
a planetesimal and will flow around a planetary embryo, although clearly 
a careful calculation of the trajectories is required to determine their fate.

We are also motivated to calculate the trajectories of 
particles carefully to determine if they will melt and cool in a manner consistent
with chondrules.
The maximum heating of the chondrules is set by their velocity relative
to the gas after passing through the shock front, which sets their
degree of melting.
Their cooling after melting is set by the optical depth between the
particles and the unshocked gas, which in turn is set by the geometry 
and spatial distribution of the particles.
These must be calculated numerically. 

We have computed the trajectories of particles by  integrating
the particles' motions through the gas, assuming the gas
flow around the planetary body is in a steady state.  
We assume the particles are tracers only, with no dynamical effect on the gas.  
Particles are accelerated by the gravity of the planetary embryo 
(with mass $M_{\rm p}$) and by the drag force due to gas of density $\rho_{\rm g}$:
\begin{equation}
m_{\rm c} \frac{d \vvc}{dt} 
 = -\frac{G M_{\rm p} m_{\rm c}}{r^3} \, \rvec 
   -\pi a_{\rm c}^{2} \, \left( \frac{ C_{\rm D} }{2} \right) \,
  \rho_{\rm g} \, 
  \left| \vvc - \vvg \right| \, \left( \vvc - \vvg \right), 
\end{equation}
where $\rvec$ is the position vector of the chondrule (with respect to the 
planet center), and $\vvc$ and $\vvg$ are the chondrule and gas velocities. 
Particles are in the Epstein limit (much smaller than the mean free
paths of gas molecules), and the drag force is given by 
\begin{equation}
C_{\rm D} = \frac{2}{3 s} \left( \frac{\pi T_{\rm c}}{T_{\rm g}} \right)^{1/2}
 + \frac{2 s^2 + 1}{\pi^{1/2} s^3} \, \exp \left( -s^2 \right)
 + \frac{4 s^4 + 4 s^2 - 1}{2 s^4} \, {\rm erf}(s),
\end{equation}
where we will assume the particle temperature $T_{\rm c}$ locally equals the 
gas temperature $T_{\rm g}$, and where 
\begin{equation} 
s = \frac{ \left| \vvc - \vvg \right| }{ (2 k T_{\rm g} / \bar{m})^{1/2} },
\end{equation} 
where $\bar{m} = 2.33 m_{\rm H}$ is the mean molecular weight of the gas
and $k$ is Boltzmann's constant (Probstein 1968).
Gas velocities $\vvg$, temperatures $T_{\rm g}$ and densities $\rho_{\rm g}$ 
are found from the output of the hydrodynamic simulations, using bilinear 
interpolation on a grid with spatial resolution of 32 km.
Particle velocities are integrated using a simple explicit scheme that is 
first-order in time, with timesteps of 0.05 s.
Particles move no more than 0.4 km during each timestep, which is significantly 
smaller than the spatial resolution of the hydrodynamics simulations. 

The trajectories of chondrules around both the planetesimal and the planetary embryo
are depicted in Figures~\ref{fig:streamlines}a and~\ref{fig:streamlines}b. 
These are seen to be significantly affected by the size of the region between the shock 
front and the body.
At the shock front the gas undergoes a nearly instantaneous shift in velocity.
Chondrules regain dynamical equilibrium with the gas only after they move through
the gas by a stopping length $V_{\rm rel} t_{\rm stop}$, where $t_{\rm stop}$ is
the stopping time (Equation 2) and $V_{\rm rel}$ is the velocity of a chondrule 
with respect to the gas. 
This quantity depends only on the gas density $\rho_{\rm g}$, not the radius of 
the planetary body. 
For both the planetesimal and planetary embryo cases, the stopping time
$t_{\rm stop} \approx \, 50 \, {\rm s}$ 
(assuming a gas density $\rho_{\rm g} = 1 \times 10^{-9} \, {\rm cm}^{-3}$),
over which time the chondrules slow from a velocity (relative to the planet)
of $8 \, {\rm km} \, {\rm s}^{-1}$ to about $2 \, {\rm km} \, {\rm s}^{-1}$.
Assuming an average velocity $\approx 5 \, {\rm km} \, {\rm s}^{-1}$, chondrules
will typically move about 250 km past the shock front. 
This stopping length is to be compared to the size of the post-shock region,
specifically the standoff distance between the shock front and the planetary 
body itself. 
For the $R_{\rm p} = 500 \, {\rm km}$ planetesimal case, this standoff distance between 
the shock front and the planetesimal is $\approx 110 \, {\rm km} \approx 0.2 \, R_{\rm p}$.  
It is therefore not surprising that all chondrules with impact parameter 
$b < 460 \, {\rm km}$ are accreted onto the planetesimal.
In contrast, for the $R_{\rm p} = 2720 \, {\rm km}$ planetary embryo case, the 
standoff distance to the planet is $\approx 2000 \, {\rm km} \approx 0.8 \, R_{\rm p}$,
or $\approx 1400 \, {\rm km} \approx 0.5 \, R_{\rm p}$ to the atmosphere, much greater than 
the stopping length.
Thus, the vast majority of chondrules dynamically recouple with the gas before hitting
the planet.  Only those chondrules with $b \ltsimeq 400 \, {\rm km}$ might be accreted, 
and chondrules that escape accretion flow around the body.
Along the way, their trajectories are affected by the KH instabilities between
the shocked gas and the primary atmosphere.  
This makes it difficult to predict their {\it exact} trajectories, but it is clear that 
they will escape accretion onto the body.

Moving forward it is important to distinguish between particles that pass through
the shock front more directly and are melted as chondrules, and those particles that
pass through the shock front far from the body and are not melted as chondrules. 
Only particles that pass through the shock front with sufficient velocity $V_{\rm rel}$
relative to the gas will melt as chondrules. 
The heating rate of chondrules after they pass through the shock front scales as
$V_{\rm rel}^{3}$ (up to the liquidus temperature). 
As we discuss in \S 4, a shock speed of $6 \, {\rm km} \, {\rm s}^{-1}$ 
is required to ensure melting of chondrules, which is equivalent to  
$V_{\rm rel} \gtsimeq 5 \, {\rm km} \, {\rm s}^{-1}$. 
The relative velocity between chondrules and gas is a function of the impact
parameter $b$, and is easily calculated for the case $b = 0$ km (chondrules and
gas impact the body directly).
Just past the shock front, chondrules retain their velocity but gas is immediately 
slowed by a factor of $(\gamma+1)/(\gamma-1) \approx 6$.
The velocity difference between gas and chondrules is thus 
$\approx (5/6) (8 \, {\rm km} \, {\rm s}^{-1}$) $\approx 6.7 \, {\rm km} \, {\rm s}^{-1}$.
As the impact parameter $b$ is increased, this relative velocity decreases.
This is because only the component of the velocity normal to the shock front is
reduced, whereas the tangential component of velocity is conserved.
As $b$ increases and the shock becomes increasingly oblique, the relative velocity 
is decreased, and therefore the chondrule heating rate is decreased. 
We therefore anticipate a maximum impact parameter for precursors to be melted as
chondrules. 

This variation of relative velocity with impact parameter is indeed seen in our 
numerical simulations, as illustrated in Figures~\ref{fig:vel}a (planetesimal case)
and~\ref{fig:vel}b (planetary embryo case).
In both cases, the maximum relative velocity at $b = 0 \, {\rm km}$ approaches
$\approx 6.7 \, {\rm km} \, {\rm s}^{-1}$, and decreases with increasing impact parameter.
In the $R_{\rm p} = 500 \, {\rm km}$ case, the relative velocity drops below
the critical threshold for melting, $5 \, {\rm km} \, {\rm s}^{-1}$, for impact 
parameters $\gtsimeq 400 \, {\rm km}$.
For the $R_{\rm p} = 2720 \, {\rm km}$ planetary case, on the other hand, 
particles experience $V_{\rm rel} > 5 \, {\rm km} \, {\rm s}^{-1}$
for impact parameters as high as $b \approx 4300 \, {\rm km}$.
If the planetary embryo had even higher speed with respect to the gas, this 
maximum impact parameter would be expanded.
Because of the slightly irregular shape of the shock front, some regions are 
not as oblique as \sout{in} others.
In the snapshot of Figure~\ref{fig:streamlines}b, for example, 
particles with $b \approx 6000 \, {\rm km}$ encounter the shock more directly
and \sout{also} may experience thermal histories consistent with chondrules.
Note that even particles with $b = 10,000$ km will be significantly heated
(by a shock with speed $2.5 \, {\rm km} \, {\rm s}^{-1}$), but will not be melted
as chondrules. 
Comparing the cross sections, the vast majority ($> 80\%)$ of particles encountered 
by the bow shock will be heated, but not melted. 

There is generally a maximum impact parameter for which particles are melted
as chondrules, but a minimum impact parameter for which particles escape accretion.
For a planetesimal with radius $R_{\rm p} = 500 \, {\rm km}$, only particles with
$b < 400 \, {\rm km}$ are heated strongly enough to melt as chondrules.  However,
only particles with $b > 460 \, {\rm km}$ escape accretion onto the planetesimal.
From this we see that {\it all} particles melted as chondrules in a planetesimal
bow shock are accreted onto the planetesimal.
For a planetary embryo with radius $R_{\rm p} = 2720 \, {\rm km}$, on the other hand, 
all particles with $b < 4300 \, {\rm km}$ (possibly up to 6000 km) are melted as
chondrules, whereas only particles with $b < 400 \, {\rm km}$ are possibly 
(but not necessarily) accreted. 
More than $99\%$ of all particles melted as chondrules in a planetary embryo
bow shock escape accretion by the body. 
Larger bodies will present a larger shock front to incoming particles, increasing
the impact parameter out to which chondrules form, and will produce shocks that
stand off farther from the body, decreasing the fraction of particles accreted.

One important caveat to the results stated above is that the chondrule trajectories 
were calculated under the assumption that the gas flow was in steady state.  
This assumption is valid for gas as it flows to the shock, and even in the immediate 
post-shock region, but is manifestly not valid at the boundary where the post-shock 
gas flows past the planetary atmosphere, since this boundary is strongly affected by KH rolls. 
Figure~\ref{fig:streamlines}b makes clear that where chondrules intercept a KH roll 
(e.g., at impact parameters 2000 km and 5000 km), the gas flow drives chondrules out of 
the KH roll.
Likewise, at impact parameter $\approx 4000$ km, characterized by the absence of KH rolls, 
the gas is flowing from the shock front inward, driving chondrules into the void. 
Thus, there is a tendency for chondrules to ``skim" the boundary surface between the 
atmosphere and the post-shock gas, although it is difficult to confirm this tendency
under the assumption of a steady-state flow.  
Chondrules in the vicinity of the flow-atmosphere boundary have a range 
of velocities $\approx 1.8 - 5.6 \, {\rm km} \, {\rm s}^{-1}$ (see below), while the 
KH rolls themselves have velocities (pattern speeds) on the order of 
$3.5 \, {\rm km} \, {\rm s}^{-1}$, so that chondrules will probably move in and out 
of KH rolls, but not at the rates they do so in this calculation, in which the gas flow 
is kept in steady state. 

The trajectories of chondrules determine their density in the chondrule-forming region.
In what follows, we specifically focus on particles that achieved temperatures high
enough to melt as chondrules.
To be specific, we consider the particles processed by the planetary embryo bow shock, 
in Figure 2b, which will pass through a slice with $Y = +3000 \, {\rm km}$ near the shock front, 
with $X \approx 7000 - 10000 \, {\rm km}$.
Considering just the subset of those particles that melted as chondrules, they have 
impact parameters $b = 400 - 4300 \, {\rm km}$, and pass through
$X = 7180 - 8060 \, {\rm km}$. 
Particles with the lower end of impact parameter ($b = 400 \, {\rm km}$) move most 
slowly, at speeds $\approx 1.8 \, {\rm km} \, {\rm s}^{-1}$, and take 2-4 hours
to reach this region.
Chondrules with the higher impact parameter ($b = 4300 \, {\rm km}$) tend to have
speeds $\approx 5.6 \, {\rm km} \, {\rm s}^{-1}$ and take just under 1 hour to reach
this region.
On average, chondrules' velocities are reduced from their pre-shock values 
($8 \, {\rm km} \, {\rm s}^{-1}$) by factors 
$\approx 2.3$, 
implying that chondrule densities will increase by a factor 
2.3 as they slow.
In addition, the chondrules are also seen to be geometrically concentrated as they
are funneled into a smaller conical volume \sout{in} past the bow shock.
This increases their density by a factor $(4300^2 - 400^2) / (8060^2 - 7180^2) = 1.37$.
Combined, the chondrule density is raised above the background value by a factor of 
3.2. 
A similar calculation for those particles with impact parameters 
$4300 \, {\rm km} < b < 10000 \, {\rm km}$ shows the density of these particles
is increased by 
a similar factor. 
These particles will not have achieved the same peak temperatures as those particles
with smaller impact parameters, but they will be almost as hot.
For example, if particles with $b = 0$ km achieved peak temperatures of 2000 K, and
peak temperature scales as $V_{\rm rel}^{3/4}$, then Figure 3b suggests that particles
with $b = 8000 \, {\rm km}$ will have peak temperatures $\approx 1400 \, {\rm K}$. 

The chondrule densities derived above bear on the frequency of compound chondrules.
The mean free path of a chondrule in a sea of chondrules is
$l_{\rm mfp} = (n_{\rm c} \, 4 \pi a_{\rm c}^{2} )^{-1}$. 
For $a_{\rm c} = 300 \, \mu{\rm m}$ and assuming a compression of a factor of 3.2 in the 
post-shock region, we find 
$l_{\rm mfp} \approx 2.7 \times 10^{4} \, {\cal C}^{-1} \, {\rm km}$.
For ${\cal C} = 10$, the mean free path is 2700 km (about the radius of the planet). 
Within the part of the flow containing chondrules, their velocities range between 1.8 
and $5.6 \, {\rm km} \, {\rm s}^{-1}$ over a distance 800 km, shorter than the mean 
free path.  
Chondrules therefore will collide with other chondrules at relative velocities $\Delta V$ 
up to and above that needed to shatter chondrules, which has been estimated to lie in the
range $1 \, {\rm m} \, {\rm s}^{-1}$ (Kring 1991) to $100 \, {\rm m} \, {\rm s}^{-1}$
(Gooding \& Keil 1981).
We assume a median threshold for shattering of $10 \, {\rm m} \, {\rm s}^{-1}$, assert
that collisions with chondrules up to (and beyond) this speed will occur, and that these
will dominate the production of compound chondrules.
The compound chondrule frequency $f$ we predict is 
\begin{equation} 
f = (\Delta V) t_{\rm plast} l_{\rm mfp}^{-1} 
= 0.04 \,  
  \left( \frac{\Delta V}{10 \, {\rm m} \, {\rm s}^{-1}} \right) \,
  \left( \frac{ t_{\rm plast} }{ 10^4 \, {\rm s} } \right) \, 
  \left( \frac{ {\cal C} }{ 10 } \right), 
\end{equation}
where $t_{\rm plast}$ is the time for which chondrules remain plastic enough to 
stick after a collision. 
The observed frequency of compound chondrules is reproduced in a planetary embryo
bow shock assuming chondrules are concentrated by factors ${\cal C} \approx 10$,
although the exact percentage will depend on the maximum speed at which chondrules
can collide and stick without shattering. 

Finally, the density of chondrules in the post-shock region will also bear on the
optical depths between chondrules and the cool, unshocked gas, which we estimate as
follows. 
The optical depth is $\tau = n_{\rm c} \, (\epsilon \pi a_{\rm c}^{2}) \, l$, where 
$l$ is the distance between the locations of chondrules and the bow shock, and 
$\epsilon \, \pi a_{\rm c}^{2}$ is the wavelength-integrated absorption cross-section, with 
$\epsilon \approx 0.8$ in the near-infrared (Li \& Greenberg 1997). 
Dust evaporates at the shock front, so the opacity is due solely to chondrules 
(Morris \& Desch 2010).
The number density of chondrules in the background nebular gas is 
\begin{equation}
n_{\rm c,0} = \frac{ (\rho_{\rm c} / \rho_{\rm g}) \rho_{\rm g} }
       { 4\pi \rho_{\rm s} a_{\rm c}^3 / 3} \approx 1 \times 10^{-8} \, {\cal C} \, {\rm cm}^{-3},
\end{equation} 
where we have assumed a chondrule radius $a_{\rm c} = 300 \, \mu{\rm m}$, an internal density
$\rho_{\rm s} = 3.3 \, {\rm g} \, {\rm cm}^{-3}$, and defined 
${\cal C} \equiv (\rho_{\rm c} / \rho_{\rm g}) / (3.75 \times 10^{-3})$. 
We estimate $l$ by considering those chondrules (objects with $b < 4300 \, {\rm km}$)
with most probable impact parameter, $b \approx 3000 \, {\rm km}$.
These chondrules pass $Y = +3000 \, {\rm km}$ at $X \approx 7800 \, {\rm km}$.
The particles with the largest impact parameter to be chondrules ($b = 4300 \, {\rm km}$)
pass through $Y = +3000 \, {\rm km}$ at $X = 8060 \, {\rm km}$.
Importantly, though, even particles with $b = 8000 \, {\rm km}$ will achieve 
near-chondrule-like temperatures of 1400 K, and pass through $Y = +3000 \, {\rm km}$
at about $X = 9000 \, {\rm km}$. 
The unshocked gas lies at $X > 10000 \, {\rm km}$ at $Y = +3000 \, {\rm km}$.
Thus the typical chondrule lies 250 km from the edge of the cloud of particles
shocked enough to melt as chondrules, but 2100 km from the bow shock
(after correcting for the tilt of shock front).
We assume chondrules lie 2200 km from the bow shock and consider 
chondrules to be concentrated above their background value by a factor of 3.2, to find
\begin{equation}
\tau = 0.16 \, \left( \frac{ {\cal C} }{10} \right) \,
               \left( \frac{ l }{ 2200 \, {\rm km} } \right).
\end{equation} 
The gradual spatial transition from shocked chondrules to unshocked gas makes it 
difficult to define $l$ more exactly. 
Nevertheless, this calculation demonstrates that if chondrules are concentrated 
by factors of about 10, then $\tau \sim 0.16$ for most chondrules melted by planetary 
embryo bow shocks.

\section{Thermal Histories of Chondrules in Planetary Embryo Bow Shocks}

Having calculated the trajectories of chondrules past the planetary embryo, an
important goal is to calculate their thermal histories.
These are influenced strongly by the absorption of infrared radiation emitted by
nearby chondrules, and by the proximity of a cooler region into which the chondrules
can radiate (without receiving a return of radiation).
In the previously studied cases of large-scale shocks driven by disk-wide gravitational
instabilities (e.g., Morris \& Desch 2010), the shock front is planar and the 
geometry is 1-D.  
The cooler region into which chondrules radiate is the pre-shock gas (ahead of a 
radiation front). 
Chondrule trajectories are parallel to each other and normal to the shock front, 
and chondrule properties do not vary in the lateral direction.
In contrast, following melting by a bow shock around a planetesimal or planetary embryo, 
the cooler region to which chondrules radiate is in the lateral direction, and the
shock clearly has a 2-D geometry.
The only correct way to calculate chondrule thermal histories is
to compute the fully 2-D radiation field in cylindrical geometry,
but the chondrule trajectories displayed in Figure~\ref{fig:streamlines}b suggest
that a perturbation to the 1-D chondrule thermal histories may yield an adequate solution. 

First, the particles are seen to flow in a more or less laminar fashion, without crossing
paths.  This suggests that chondrule thermal histories are a function of distance past
the shock front, as in the 1-D example.
Second, chondrules on parallel trajectories do not differ from each other significantly.
As stated above, it takes about 3 hours for chondrules with the lowest initial impact 
parameter ($b = 400 \, {\rm km}$), and about 1 hour for chondrules with the highest impact
parameter ($b = 4300 \, {\rm km}$), to reach a certain representative point downstream 
($Y = +3000 \, {\rm km}$). 
We show below that typical chondrule cooling rates are roughly $60 \, {\rm K} \, {\rm hr}^{-1}$,
so that at this location all chondrules are within about 100 K of each other.
Compared to the typical chondrule temperature at this location ($\approx 1600 \, {\rm K}$),
this difference is not significant.
Third, these properties hold even as the particles move downstream by roughly 20,000 km
(not shown in Figure~\ref{fig:streamlines}b). 
Further downstream, rarefaction of the shocked gas is presumed to take place.
Indeed, some hints of this are seen in the slightly diverging chondrule trajectories 
in Figure~\ref{fig:streamlines}b, but this does not affect the gross geometry.
Certainly the flow remains basically the same for at least the 5 hours it takes
for chondrules in our simulations to cool below their solidus temperatures.
These facts suggest that a simple alteration to the 1-D code of Morris \& Desch (2010), to 
allow for loss of energy by radiation into the unshocked gas, may allow a useful estimate
of chondrule thermal histories.

We accommodate for radiative losses in an approximate way using the formulation of Morris et al.\
(2010a,b). 
Specifically, we conduct a 1-D simulation as in Morris \& Desch (2010), based on Desch \&
Connolly (2002), including emission, absorption, and transfer of continuum radiation from 
solids, and the molecular radiation from ${\rm H}_{2}{\rm O}$, 
dissociation and recombination of ${\rm H}_{2}$ molecules, and evaporation of small particles. 
However, we replace the 1-D radiation field $J$ seen by chondrules with one that is a mixture
of the local 1-D radiation field and the radiation field they see from the cool gas on the other 
side of the bow shock.
To calculate this mixed radiation field, we consider chondrules to lie in one of two 
semi-infinite spaces separated by a plane, which represents the bow shock.
As seen in Figure~\ref{fig:streamlines}b, chondrules tend to move very close to, and parallel
to, the shock front. 
We assume they all remain at a  fixed optical depth $\tau$ from the bow shock.
On the unshocked side of the bow shock, we assume the temperature is that of the 
background nebula, $T_{\rm bkgrnd} \approx 150 \, {\rm K}$. 
On the other side lies heated chondrules, and the hot ($T \sim 10^{3} \, {\rm K}$)
gas in the planetary embryo's wake.  
The column density across the wake region is $\sim 10 \, {\rm g} \, {\rm cm}^{-2}$, 
which for a solar-composition gas yields a column density of water molecules 
$N_{\rm H2O} \sim 10^{21} \, {\rm cm}^{-2}$, which is optically thick in the infrared
(Morris et al.\ 2009). 
In addition, gas that might contain dust is mixed into the wake region, especially at 
$Y > +8000 \, {\rm km}$.
This justifies the assumption of a semi-infinite space at high temperature, which
we assume is the local temperature of chondrules, $T_{\rm ch}$. 
Making these assumptions, it is straightforward to show that chondrules see a radiation 
field with mean intensity
\begin{equation}
J(\tau) = B(T_{\rm ch}) + \frac{1}{2} \left[ B(T_{\rm bkgrnd}) - B(T_{\rm ch}) \right] \, {\rm E}_{2}(\tau),
\end{equation} 
where $B$ is the Planck radiation function and ${\rm E}_{2}$ is the second exponential integral.

We replace the radiation field seen by chondrules with this quantity, where $\tau$ is an input 
parameter describing the optical depth between the chondrules and the gas.
We note that the previous work of Morris et al.\ (2010a,b) differed
in two ways.  In that work, ${\rm E}_{2}(\tau)$ was approximated by
$\exp(-\tau)$, which is a reasonably good assumption.  That work also
assumed the wake of the planetesimal was {\it not} optically thick,
so that $\exp(-\tau)$ was used instead of the more appropriate
${\rm E}_{2}(\tau)$ used here.  Because of these differences, cooling
rates are roughly an order of magnitude slower here than found by
Morris et al.\ (2010a,b).  As discussed above in \S 3, we consider $\tau = 0.16$ to be a typical value for those particles
that melt as chondrules (impact parameters from 400 to 4300 km), provided ${\cal C} \approx 10$.  
Based on this, ${\rm E}_{2}(0.16) / 2  = 0.313 = \exp(-1.16)$, meaning that the radiation field
seen by chondrules is roughly 69\% the radiation field they would normally see in the 1-D shock,
mixed with 31\% of the background radiation field. 
Chondrules are warmed by the radiation field from nearby chondrules, but not as much as they
would be in a 1-D shock.

We now consider the possible effects of H$_2$O line cooling in this case.  Morris \& Desch (2010) found that molecular line cooling due to H$_2$O was negligible in large scale shocks, due to the combined effects of high column densities of water seconds after the shock, backwarming, and, most importantly, ${\rm H}_{2}$ recombinations.  However, we must consider whether line cooling would become significant in the geometry of a bow shock.  Recall from above that the column density of water molecules across the wake region is $N_{\rm H2O} \sim 10^{21} \, {\rm cm}^{-2}$, so the region is optically thick to line photons shortly after the shock (Morris et al.\ 2009).  Not only is the wake region optically thick to line photons, but the high column density of water also tends to block continuum
radiation escaping through the wake region.  Although the effects of backwarming may be reduced somewhat due to the bow shock geometry, buffering by ${\rm H}_{2}$ recombinations will be unaffected.  The combination of high column densities of water and buffering by ${\rm H}_{2}$ recombinations will dominate over any reduction in the efficiency of backwarming due to the geometry.  Therefore, we still consider H$_2$O line cooling negligible in bow shocks. 

In Figure 5, the thermal histories of chondrules are shown for four possible values of the shock speed: 
$V_{\rm s} = 5 \, {\rm km} \, {\rm s}^{-1}$, $6 \, {\rm km} \, {\rm s}^{-1}$, 
$7 \, {\rm km} \, {\rm s}^{-1}$ and $8 \, {\rm km} \, {\rm s}^{-1}$.
In each case chondrules are heated by absorbing infrared radiation emitted by other chondrules
in the vicinity, by thermal exchange with the gas in the post-shock region, and by frictional
heating in the minute or so past the shock where the chondrules move at supersonic speeds
$V_{\rm rel}$, with respect to the gas.
As described above, the radiation field seen by chondrules has been artifically lowered from 
the radiation field they would normally see, using the factor $\tau = 0.1$ ($\tau' = 1$). 
The peak temperature is sensitive to the supersonic heating rate, which scales as $V_{\rm rel}^{3}$ (up to the liquidus temperature),
and strongly correlates with shock speed. 
We find 
$T_{\rm peak} \approx 1430 \, {\rm K}$ for $V_{\rm s} = 5 \, {\rm km} \, {\rm s}^{-1}$, 
$T_{\rm peak} \approx 1800 \, {\rm K}$ for $V_{\rm s} = 6 \, {\rm km} \, {\rm s}^{-1}$, 
$T_{\rm peak} \approx 1820 \, {\rm K}$ for $V_{\rm s} = 7 \, {\rm km} \, {\rm s}^{-1}$, and
$T_{\rm peak} \approx 1820 \, {\rm K}$ for $V_{\rm s} = 8 \, {\rm km} \, {\rm s}^{-1}$.
Note that thermal histories become increasingly uncertain for times $> 5$ hours
(i.e., for temperatures $< 1400$ K) because the assumption of a 1-D geometry becomes increasingly
less valid. 

Chondrule peak temperatures must exceed 1770 - 2120 K to produce porphyritic textures
(Hewins \& Connolly 1996; Desch \& Connolly 2002).
Formation of chondrules therefore requires $V_{\rm s} \gtsimeq 6 \, {\rm km} \, {\rm s}^{-1}$,
or $V_{\rm rel} \gtsimeq 5 \, {\rm km} \, {\rm s}^{-1}$. 
For the runs where peak temperatures were sufficient to melt the chondrules, the 
cooling rates through the crystallization temperature range ($\approx 1400 - 1820 \, {\rm K}$,
Desch \& Connolly 2002) were $< 10^2 \, {\rm K} \, {\rm hr}^{-1}$ (the cooling rate for
chondrules in the shock with $V_{\rm s} = 6 \, {\rm km} \, {\rm s}^{-1}$ did not begin to
cool slowly until their temperatures dropped below $1400 \, {\rm K}$).
If we repeat our calculations with ${\cal C} \gtsimeq 100$, we recover the  optically thick,
1-D limit for which cooling rates are $< 10^2 \, {\rm K} \, {\rm hr}^{-1}$. 
If we reduce ${\cal C}$ below 10, the chondrule cooling rates increase.
They may not increase significantly if the wake region remains optically thick, but if it does
not, then ${\cal C} \sim 1$ leads to cooling rates $\sim 10^{3} \, {\rm K} \, {\rm hr}^{-1}$
(Morris et al.\ 2009). 
From all this, we conclude that both peak temperatures and cooling rates are consistent with
formation of porphyritic textures of chondrules, provided shocks speeds are 
$V_{\rm s} > 6 \, {\rm km} \, {\rm s}^{-1}$ ($V_{\rm rel} > 5 \, {\rm km} \, {\rm s}^{-1}$),
and ${\cal C} > 10$. 
Lower particle concentrations may also be consistent with cooling rates 
$< 10^3 \, {\rm K} \, {\rm hr}^{-1}$, although we have not tested this.
The required shock speeds are obtained so long as the planetary embryo has sufficient 
orbital eccentricity and/or inclination.  

Because cooling rates are sensitive to chondrule concentrations ${\cal C}$, it is worth
considering likely values of ${\cal C}$ as well as the lengthscales over which such concentrations are achieved.
One mechanism that may lead to a high concentration of chondrules is settling 
to the midplane, which is inferred observationally in certain young 
protoplanetary disks (Furlan et al.\ 2011).
In this scenario, the level of turbulence at the midplane is low enough that
chondrules have a scale height $h_{\rm c}$ significantly smaller than the scale 
height $H$ of the gas.
Dubrulle et al.\ (1995) calculated the scale height of particles as a function
of the level of turbulence, as parameterized by $\alpha$.
This can be written in terms of ${\cal S} = \alpha / (\Omega \; t_{\rm stop})$, 
where $\Omega$ is the Keplerian orbital frequency, 
$t_{\rm stop} = \rho_{\rm s} a_{\rm c} / \rho_{\rm g} c_s$ is the chondrules'
aerodynamic stopping time as defined in Section 3, $\nu = \alpha c_s H$ is the 
turbulent viscosity, $c_s$ is the sound speed of the gas, and $H = c_s / \Omega$.
The chondrule scale height is 
\begin{equation}
h_{\rm c} = h \left[ 1 + \left( \frac{h}{H} \right)^2 \right]^{-1/2},
\end{equation}
where $h / H = (\gamma+1)^{-1/4} {\cal S}^{1/2}$.
If $\alpha \ll \Omega \; t_{\rm stop}$, then $h \ll H$ and 
chondrules may be concentrated at the disk midplane. 
Assuming $\rho_{\rm g} = 1 \times 10^{-9} \, {\rm g} \, {\rm cm}^{-3}$,
$C = 0.7 \, {\rm km} \, {\rm s}^{-1}$ ($T = 150 \, {\rm K}$) and 
$\Omega = 7 \times 10^{-8} \, {\rm s}^{-1}$ ($r = 2 \, {\rm AU}$),
$t_{\rm stop} = 1400 \, {\rm s}$ and $\Omega \; t_{\rm stop} = 1 \times 10^{-4}$.
If $\alpha \ll 10^{-4}$ then chondrules will concentrate to the midplane.

If the turbulent viscosity of the disk is dominated by the magnetorotational
instability, it is possible, but not certain, that $\alpha$ may be low enough at 
the midplane to allow chondrules to settle. 
Even low values of $\alpha$ are insufficient to allow chondrules to settle: 
for $\alpha = 1 \times 10^{-4}$, ${\cal S} = 1$, $h / H = 0.8$ (for $\gamma = 1.4$) 
and $h_{\rm c} = 0.8 H$, signifying relatively good mixing of chondrules with the gas.
To obtain ${\cal C} = 10$
requires $\alpha < 1.6 \times 10^{-6}$, so that ${\cal S} < 1.6 \times 10^{-2}$,
$h / H < 0.101$ and $h_{\rm c} < 0.100 H$, implying significant settling to the midplane.
Chondrule concentrations would then be 10 times their background nebular values, 
over a thickness $\ltsimeq H / 10$.

This analysis is complicated by the possibility that a significant fraction of the
solids mass may be locked up in large planetesimals.
If half of the solids mass is locked up in planetesimals, the same optical depths
in the chondrule-forming region require chondrules to be concentrated by a factor of 
20 instead of 10, which would require $\alpha < 3.9 \times 10^{-7}$ so that 
$h / H < 2.5 \times 10^{-3}$ and $h_{\rm c} < 0.050 H$.
If 90\% of the solids mass is locked up in planetesimals, the same optical depths 
would require a concentration by a factor of 100 instead of 10, which would require 
$\alpha < 1.6 \times 10^{-8}$ so that $h_{\rm c} < 0.010 H$.
It is not clear what value of $\alpha$ is appropriate at the midplanes of protoplanetary
disks in which the magnetorotational instability cannot act at the midplane.
In principle, lower values are possible (e.g., Turner et al.\ 2010), but values 
$\alpha \approx 10^{-4}$ are thought to be appropriate (Fleming \& Stone 2003),
which would make it difficult for significant concentrations of chondrules to be 
achieved. 


Turbulence itself provides a second mechanism for concentrating chondrules. 
Cuzzi et al.\ (2001) demonstrated that particles with an aerodynamic stopping 
time equal to the turnover time of the Kolmogorov-scale eddies, can be concentrated 
in the stagnant zones between turbulent eddies.
Particles with this stopping time have a size
\begin{equation} 
a_{\rm c} = \frac{ \rho_{\rm g} c_s \nu_{\rm m}^{1/2}}
                              { \rho_{\rm s} \Omega (\alpha c_s H)^{1/2} }.
\end{equation}
The molecular viscosity $\nu_{\rm m} = \eta / \rho_{\rm g}$, where
$\eta$ is given by Sutherland's formula.  
Following the discussion in Desch (2007), and assuming a solar composition 
gas with $T = 150 \, {\rm K}$ and the parameters above, we calculate
$\eta = 59 \, \mu{\rm P}$ and $\nu_{\rm m} = 5.9 \times 10^{4} \, {\rm cm}^2 \, {\rm s}^{-1}$.
For $\alpha = 10^{-4}$ we determine the size of particles necessary for concentration
is $a_{\rm c} = 280 \, \mu{\rm m}$.
Chondrule precursors with radius $300 \, \mu{\rm m}$, as we have assumed in our model, 
have almost the exact size to be optimally concentrated.
Particle concentrations ${\cal C} \sim 10$ are easily achieved (Cuzzi \& Hogan 2003). 

Even if a significant fraction of the mass of solids is locked up in large planetesimals,
there will be many regions in the nebula that have these concentrations of chondrule precursors.
For example, in nebular gas with $\alpha = 10^{-4}$, nearly 80\% of all chondrule precursors are
in regions where their densities are concentrated by factors of 10 above background 
values, 70\% are in regions with concentrations 20 times their background density,
and 20\% are in regions with concentrations 100 times their background density
(Cuzzi et al.\ 2001). 
Of course, only small fractions of the {\it volume} of the nebula see concentrations
this high, but these regions are where most chondrule precursors reside.
The lengthscale over which a concentration ${\cal C}$ is achieved roughly scales as
$(10^6 \, {\rm km}) / {\cal C}$ (Cuzzi \& Hogan 2003), so even regions with ${\cal C} = 100$
are $> 10^4 \, {\rm km}$ in extent and thus wider than the planetary embryo bow shock 
itself.
We are justified in assuming a uniform value of ${\cal C}$ when computing chondrule 
thermal histories.
Note that while a planetary embryo will pass through clumps with ${\cal C} = 10 - 100$,
with sizes $10^{5} - 10^{4} \, {\rm km}$, in only 3 - 0.3 hours, the newly-formed chondrules remain
in the vicinity of each other. 
To fix ideas, we assume that half of the solids mass is locked in large planetesimals,
so that the optical depths assumed in the thermal histories above requires concentrations
by factors of 20, which are achieved over regions 50,000 km in extent, which contain
70\% of all chondrule precursors .  
If the planetary embryo is on an orbit eccentric and/or inclined enough to yield shock speeds 
$> 6 \, {\rm km} \, {\rm s}^{-1}$, the chondrule precursors  in these clumps would be melted and
would cool in a manner consistent with the dominant porphyritic textures. 
 
\section{Chemical Environment of Planetary Embryo Bow Shocks}

We now consider the formation environment experienced by chondrules in bow shocks 
around planetary embryos, which we expect to be quite different from the environment 
in planetesimal bow shocks.  
Dauphas \& Pourmand (2011) have suggested that Mars accreted roughly half
its mass within $1.8^{+0.9}_{-1.0} \, {\rm Myr}$ after the formation of CAIs.
It therefore must have contained abundant live ${}^{26}{\rm Al}$ when it
formed, sufficient to melt the planet (Dauphas \& Pourmand 2011; Grimm \& 
McSween 1993).
It is presumed that such a body would also form a magma ocean, would convect,
and would rapidly outgas volatile species.
Zahnle et al.\ (2007) have considered the outgassing of volatiles from
the Earth following the Moon-forming impact, when the Earth was in its
magma ocean state, and concluded that an atmosphere in chemical equilibrium
with the magma could rapidly develop in $< 10^3 \, {\rm yr}$.
Elkins-Tanton (2008) showed that the mantle of an Earth-sized planet takes 5 Myr
to become 98\% solidified.
This suggests that planetary embryos could maintain a magma ocean and outgas volatile 
species for several Myr, perhaps aided by the continuous decay of ${}^{26}{\rm Al}$.

%
%
%
%
%
%
%

The composition of the planetary embryo's atmosphere probably was similar
to the Earth's outgassed atmosphere during its magma ocean stage.
Zahnle et al.\ (2007) considered the atmosphere outgassed from the  
proto-Earth's magma ocean.
They calculated that a fraction of the Earth's ${\rm CO}_{2}$ would be outgassed, 
equivalent to a partial pressure $P_{\rm CO2} \approx 100 \, {\rm bar}$, and that 
a substantial fraction of the ${\rm H}_{2}{\rm O}$ would be outgassed too, 
equivalent to a partial pressure $P_{\rm H2O} \approx 100 \, {\rm bar}$. 
Other moderately volatile species such as S, Na, Zn, Cl and K, 
are also expected to be outgassed during the magma ocean stage
(Zahnle et al.\ 2007; Holland 1984).

We estimate the abundances of these species in the atmosphere of the planetary
embryo by assuming that while outgassing may continue for many Myr, at any one
instant in time the magma ocean is in chemical equilibrium with the atmosphere.
The mass fraction of ${\rm CO}_{2}$ dissolved in the magma scales linearly with 
the partial pressure in the atmosphere, as 
\begin{equation}
x_{\rm CO2} = 4.4 \times 10^{-7} \, \left( \frac{ P_{\rm CO2} }{1 \, {\rm bar}} \right) 
\end{equation}  
(Stolper \& Holloway 1988). 
It is straightforward to show that if the planetary embryo has the same total
abundance of ${\rm CO}_{2}$ (C) that $P_{\rm CO2}$ will scale as $g^2$, where
$g$ is the gravitational acceleration at the planet surface. 
We accordingly estimate $P_{\rm CO2} \approx 10 \, {\rm bar}$.
The solubility behavior of ${\rm N}_{2}$ is similar to that of ${\rm CO}_{2}$ 
(Fricker \& Reynolds 1968) and we assume $P_{\rm N2} \sim 0.1 \, {\rm bar}$. 
The mass fraction of water dissolved in the magma, $x_{\rm H2O}$, is related to 
the partial pressure of water vapor in the atmosphere, $P_{\rm H2O}$, 
by the relation 
\begin{equation}
x_{\rm H2O} = 6 \times 10^{-7} \, \left( \frac{ P_{\rm H2O} }{ 1 \, {\rm dyn} \, {\rm cm}^{-2} } \right)^{0.54}
\end{equation}
(Fricker \& Reynolds 1968). 
Analyses of Martian meteorites have led to estimates of Mars's bulk water content in
a range of 140-250 ppm (McCubbin et al.\ 2011), or 0.5wt\% (Craddock \& Greeley 2009) 
to $\approx 1.4 - 1.8$wt\% (McSween \& Harvey 1993; McSween et al.\ 2001; Dann et al.\ 2001).
We take a median value, $x_{\rm H2O} = 0.2$wt\%, as a representative value
(a mass fraction equivalent to the Earth having 8 oceans, in line with current
estimates by Mottl et al.\ 2007), so that the total mass of ${\rm H}_{2}{\rm O}$ in the 
planetary embryo is $6.6 \times 10^{23} \, {\rm g}$, and $P_{\rm H2O} = 3.3 \, {\rm bar}$.
Assuming $M_{\rm p} = 3.31 \times 10^{26} \, {\rm g}$, 
$R_{\rm p} = 2720 \, {\rm km}$, and a gravitational acceleration 
$g = 300 \, {\rm cm} \, {\rm s}^{-2}$, we determine an equilibrium mass of 
${\rm H}_{2}{\rm O}$ in the atmosphere of
$1.0 \times 10^{22} \, {\rm g}$, which is a small fraction of the total mass.
The equilibrium abundance of Na in the atmosphere is complicated by 
the fact that it may dissolve in the magma either as ${\rm Na}_{2}{\rm O}$ 
or as ${\rm NaOH}$, in which case its solubility is affected by the water 
content.
van Limpt et al.\ (2006) considered Na solubility for silicate glasses and 
found the following relationship to hold when Na dissolves as NaOH (i.e. in 
the presence of H$_2$O): 
\begin{equation}
P_{\rm NaOH} \approx 1.2 \times 10^{-3} \, \left( \frac{ P_{\rm H2O} }{ 1 {\rm bar} } \right)^{0.5} \, {\rm bar}.
\end{equation}
For $P_{\rm H2O} = 3.3 \, {\rm bar}$, we find $P_{\rm NaOH} \approx 2 \times 10^{-3} \, {\rm bar}$.
Altogther, we infer that if the bulk abundance of water on the planetary embryo is 0.2\%, then 
the planetary embryo's atmosphere during the magma ocean stage is mostly (88wt\%) ${\rm CO}_{2}$ 
with partial pressure 10 bar, some (12wt\%)  ${\rm H}_{2}{\rm O}$ with partial pressure 3.3 bar, 
and alkalis and volatiles being trace species, with Na having a partial presure 
$\approx 2 \times 10^{-3} \, {\rm bar}$. 
The total mass of the atmosphere is $\approx 4 \times 10^{22} \, {\rm g}$. 

This atmosphere will be continuously stripped by KH instabilities at
the interface between the atmosphere and the post-shock gas streaming by the planet.
We did not include such an outgassed atmosphere, but the primary atmosphere accreted
by the planetary embryo in our simulations plays the same role.
The stripping by KH rolls at the interface between the planetary atmosphere and the 
post-shock gas, in particular, is expected to follow the same general trends. 
We find using dye calculations (see \S 2) that the atmosphere is lost at a rate 
$\approx 2 \times 10^{12} \, {\rm g} \, {\rm s}^{-1}$,
about 1\% of the rate at which the planetary embryo intercepts nebular gas.
This is in line with similar rates of KH stripping in other examples of strongly bound gas
(e.g., around protoplanetary disks; Ouellette et al.\ 2007).
This is sufficiently rapid to deplete the atmosphere after about 600 years, but 
outgassing can continue to replenish the atmosphere as the magma ocean convects 
(on shorter timescales) and bring new volatiles to the surface. 
The planetary inventory of water, in particular, in this example could persist for 
$8 \times 10^4 \, {\rm yr}$.  The reservoirs of Na and other species would persist for
many Myr.

Not only are gases from the planetary embryo's atmosphere stripped by KH rolls at the
interface with the shocked gas, as seen in Figure \ref{fig:streamlines}; chondrules also
make excursions in and out of these rolls and this planetary gas. 
Chondrules will be exposed, at least some of the time, to volatiles such as ${\rm H}_{2}{\rm O}$
and Na outgassed from the planetary embryo's magma ocean.
The scale height of the atmosphere proper (which has been shocked to $\approx 2500 \, {\rm K}$) 
is approximately 185 km, so even at the 800 km distance between the planetary surface and
the boundary layer at the top of the atmosphere, gas pressures should be $\sim 1\%$ of their 
values at the surface.
We estimate $P_{\rm H2O} \sim 3 \times 10^{-2} \, {\rm bar}$ and 
$P_{\rm Na} \sim 2 \times 10^{-5} \, {\rm bar}$.
Chondrules passing through KH rolls will experience approximately these partial pressures 
of volatile species. 

It is important to note that when we calculated approximate chondrule thermal histories in 
\S 4, we assumed the only gas to which chondrules were exposed was the shocked nebula gas,
with pressure $\sim 10^{-3} \, {\rm bar}$.
If chondrules pass in and out of the plumes of planetary atmosphere gas in KH rolls, they will 
see much higher presssures, $\sim 10^{-1} \ {\rm bar}$. 
This could potentially affect chondrule thermal histories in two ways: during the first minute
past the shock front, higher gas densities can more rapidly decelerate the chondrule and heat 
it by friction; and thermal exchange with the hot gas will become more efficient.
Neither of these effects is likely to drastically alter the thermal histories of the chondrules
considered here, because nearly all of the chondrule deceleration takes place in the shocked
nebula gas during the first minute, outside of the planetary atmosphere (see Figure~\ref{fig:streamlines}b).
Also, the denser plumes of gas are not likely to be at significantly different temperatures,
and chondrule temperatures will be dominated by radiative effects. 
A full treatment of chondrule thermal histories in future work must more fully consider these effects.

Elevated partial pressures of volatiles may explain puzzling features of the chondrule
formation environment. 
An enhanced abundance of water vapor is potentially consistent with indicators of elevated 
oxygen fugacity during chondrule formation (Krot et al.\ 2000; Fedkin \& Grossman 2006; 
Grossman et al.\ 2011).
Partial pressures $P_{\rm H2O} > 10^{-3} \, {\rm bar}$ have been suggested to explain the 
fayalite content of chondrule olivine (Fedkin \& Grossman 2006). 
Likewise, the high partial pressure of Na vapor may potentially explain the finding 
by Alexander et al.\ (2008) that olivine phenocrysts in chondrules from Semarkona
contain $\sim 0.010 - 0.015 {\rm wt}\%$ ${\rm Na}_{2}{\rm O}$, even at their cores, 
strongly implying that Na was dissolved in the melt as the olivine crystallized. 
Alexander et al.\ (2008) expressed the needed partial pressures in terms of a 
density of solids, assuming that the Na derived from the chondrule melts 
and that 10\% of this Na entered the gas phase. 
Their inferred chondrule densities ranged from low values up to at least
$\approx 400 \, {\rm g} \, {\rm m}^{-3}$.
Assuming the chondrule is 0.01wt\% ${\rm Na}_{2}{\rm O}$, this implies a Na vapor density 
$3 \times 10^{-9} \, {\rm g} \, {\rm cm}^{-3}$ and a partial pressure of Na 
$\approx 2 \times 10^{-5} \, {\rm bar}$.
Remarkably, the volatiles escaping from the planetary embryo's magma ocean and mixing
with the post-shock gas may provide the partial pressures of ${\rm H}_{2}{\rm O}$ and Na
assumed to have been experienced by chondrules.

We have not modeled in detail the production of a secondary atmosphere, nor quantified in detail
how that atmosphere is stripped, or how often chondrules encounter this stripped gas.
We have demonstrated that chondrules do make excursions into gas that is strongly bound to
the planet, and that this gas may contain elevated abundance of species outgassed during a 
planet's magma ocean stage, including ${\rm H}_{2}{\rm O}$, ${\rm CO}_{2}$, Na, and K vapor,
in particular.  
Our best estimates of the partial pressures of water vapor and Na vapor are consistent with
values needed to provide an environment with elevated oxygen fugacity and the ability to 
suppress evaporation of alkalis, although our results depend on the assumption $x_{\rm H2O} = 0.2$wt\%.
Further tests are required to better test these ideas quantitatively. 

\section{Masses of Chondrules Formed by Planetary Embryo Bow Shocks}

A very large number of chondrules can be produced by a single planetary embryo on 
an eccentric and/or inclined orbit.  
Chondrules will be produced as long as the velocity difference between the body 
and the gas is $V_{\rm s} \gtsimeq 6 \, {\rm km} \, {\rm s}^{-1}$, which requires
the eccentricity $e$ and inclination $i$ of the body to be sufficiently large. 
Over time, however, the interaction of a planetary embryo with nebular gas will 
cause its eccentricity and inclination to damp to low values. 
The total number of chondrules produced depends on the $e$ and $i$ damping timescales.  
In this section we construct a model for estimating the orbital conditions, duration, 
and mass processed during a single scattering event. 
The model includes $e$ and $i$ damping due to gas drag and disk torques, as well 
as semi-major axis evolution due to energy dissipation from drag.  
Migration due to torques is ignored for convenience and to reflect the fact that 
the magnitude and {\it direction} of migration can differ depending on the equation 
of state (e.g., Paardekooper \& Mellema 2006) and on whether radiative transfer is 
included or not (e.g., Bitsch \& Kley 2010). 
Below, we discuss the methods for including torques and drag, in turn, and then 
combine them to estimate the orbital evolution and mass of chondrules processed.

\subsection{Eccentricity Damping of Planetary Embryos\label{sec:ecc_damp}}

We first consider eccentricity damping by torques. 
Recent simulations by Cresswell et al.\ (2007) and Bitsch \& Kley (2010) demonstrate 
that eccentric planets in radiatively cooling disks follow analytic limits.  
When the eccentricity $e$ is high, $de/dt\vert_h = -K_e \, e^{-2}$ (Papaloizou \& Larwood 2000), 
where $K_e$ is a constant.  
When $e$ is low, $de/dt$ follows an exponential decay that is well characterized by 
$de/dt\vert_\ell = -e / t_{\rm ecc} = -e \, 0.78 \, q \, \left( \Sigma a^2/M_*\right) h^{-4} \, \Omega$, 
where $q$ is the planet-to-star mass ratio, $a$ is the planet's semi-major axis, $M_*$ is the mass of 
the star, $h=H/a$ for scaleheight $H(a)$ at heliocentric distance $a$, $\Sigma(a)$ is the total 
surface density at heliocentric distance $a$, and $\Omega(a)$ is the average angular speed of the 
planet (Tanaka \& Ward 2004).  
The low-$e$ limit is valid for $e \lesssim 0.1$.  
Using these results, we explore parameter space without using costly hydrodynamics simulations 
by making the {\it ansatz} that the total rate is the harmonic sum of the limits, such that 
$de/dt = -K_{e} e/(K_{e} t_{\rm ecc}+e^3)$.  
We select a family of solutions by varying only $K_{e}$ and assuming that $K_{e} t_{ecc}$ is 
a constant over the parameter space of interest.
Based on Cresswell et al.\ (2007) we set $K_e t_{ecc} = 0.00253$, which allows us to match 
their eccentricity evolutions.   

For the inclination evolution, the rates follow the same limiting behavior as in eccentricity damping, 
with $di/dt\vert_\ell = -i / t_{\rm inc} = -i \, 0.544 \, q \, \left( \Sigma a^2/M_*\right) h^{-4} \, \Omega$ 
(Tanaka \& Ward 2004) and $di/dt\vert_h = -K_i \, i^{-2}$ (Bitsch \& Kley 2011).  
Based on the simulations of Bitsch \& Kley (2011), we make the same assumptions for the total inclination 
damping as we have made for the eccentricity damping, and we vary $K_{i}$ while keeping constant 
$K_i t_{\rm inc} = 0.00075$.  
Strictly speaking, the inclination and eccentricity damping are coupled, but they are treated independently 
for the estimates we present here. 

For disk torques, we neglect dynamical friction 
between the embryo and
 a sea of planetesimals, which 
 is given by
 $de/dt \vert_{\rm DF} = - \ln (1+\Lambda^2) G \Sigma_p q /(\sqrt 2 \Omega a e^3)$ (Ford \& Chiang 2007). Here, $\Sigma_p$ is the surface density of planetesimals and the natural log term is the Coulomb parameter. 
The ratio $(de/dt\vert_\ell) / (de/dt\vert_{\rm DF}) \approx (e/h)^4 \Sigma/\Sigma_p (\ln(1+\Lambda^2))^{-1}$.
For typical parameters ($\Sigma/\Sigma_p$ \app 100, $\ln (1+\Lambda^2)\sim10$, and h \app 0.05)
dynamical friction
is unimportant until
$e \app 0.03$.
We show below that chondrule-producing shocks are only achieved for $e\gtrsim 0.1$. Therefore, gas disk torques will dominate over dynamical friction.

Now we consider the effects of gas drag on large solids.  
Planetesimals and planets are much larger than the mean free path of gas molecules 
$\lambda$, so the drag is in the Stokes regime, i.e., the Knudsen number 
${\rm Kn} \equiv 2 R_{\rm p} / \lambda \ll 1$.
In this regime the gas drag takes the form 
$d(\delta v) / dt = -3 \zeta \, {\rm Kn} \, k_d \, \delta v$, 
where $\delta v$ is the velocity difference,  $k_d$ is a coupling coefficient, and 
$\zeta = (8/\pi)^{1/2} \, (\rho_{\rm g} c_{\rm a}) / (\rho_{\rm s} R_{\rm p})$
$= (8\gamma / \pi)^{1/2} \, t_{\rm stop}^{-1}$,
where $c_{\rm a}$ is the adiabatic sound speed and $t_{\rm stop}$ and other 
quantitites are defined as before.
The coupling coefficient $k_d$ depends on the Reynolds number 
${\rm Re} = 3 \, (\pi/8)^{1/2} \, \mathcal{M} / {\rm Kn}$, where 
$\mathcal{M} = \delta v / c_{\rm s}$.
In the regime of interest (${\rm Re} > 1500$), $k_d = 0.11 \, {\rm Re}$ (Paardekooper 2007).  
This yields $d(\delta v) / dt  \approx -(\delta v) / t_{\rm stop}$, in accord with 
Adachi et al.\ (1976).

The evolution of the planetary embryo's orbit can now be determined if $\delta v$ can be
expressed as a function of $e$ and $i$. 
In general $\delta v$ will vary along the body's orbit, depending on the value of the
true anomaly $\phi$. 
We model the damping by solving Kepler's equation for following the orbit of a planet, and by using 
a predictor-corrector for modifying the eccentricity and inclinations over a timestep. 
The time derivatives of the radial and tangential velocity components for a Keplerian orbit are 
\begin{eqnarray}
\frac{dv_{\phi}}{dt} & = & \left(\frac{\mu}{a(1-e^2)}\right)^{1/2}
                           \left[ \left( \frac{e}{1-e^2}\left(1+e\cos{\phi}\right)+\cos\phi\right) \frac{de}{dt}
                          -e\sin\phi\frac{d \phi}{dt}\right] \\
\frac{dv_r}{dt}      & = & \left(\frac{\mu}{a(1-e^2)}\right)^{1/2}
                           \left[ \left( \frac{e^2}{1-e^2}+1\right) \frac{de}{dt}\sin\phi
                          +e\cos\phi\frac{d \phi}{dt}\right],
\end{eqnarray}
where $\mu=G(M_*+M_p)$.
We find an expression for the change in eccentricity by keeping only the terms with $de/dt$, as the other 
terms represent the unaltered Keplerian orbit.  
This gives 
\begin{equation}
\left(\frac{ d \delta v}{dt}\right)^2  =  \frac{\mu}{a(1-e^2)}\left(\frac{d e}{dt}\right)^2
 \left[\left(\frac{e}{1-e^2}\left(1+e\cos\phi\right)+\cos\phi\right)^2
      +\left(\left( \frac{e^2}{1-e^2}+1\right) \sin\phi\right)^2 \right].
\end{equation}
The change in eccentricity over time can thus be expressed in terms of the change of the
relative velocity over a computational time step.
A similar approach can be taken with the inclination, where 
$d(\delta v_z)/dt \approx (\mu / r_{\rm cyl})^{1/2} \, d i/dt$, where $r_{\rm cyl}$ is the 
projected radial separation. 
We have assumed small $i$ here because drag predominantly takes place within 
one scale height of the disk, and we in fact only include drag when $\vert z\vert < H$.
Finally, we use the energy gained or dissipated in a timestep, $\delta E$, to computed an updated
semi-major axis $a = -(\mu/2) (E + \delta E)^{-1}$.


We now apply these formulae to conditions in protoplanetary disks. 
In our orbital damping calculations we use a disk model with gas surface density 
$\Sigma(r) = 149 \, (r / 5.2 \, {\rm AU})^{-1.5} \, {\rm g} \, {\rm cm}^{-2}$, a gas scale height 
$H = 0.05 r$, a midplane density $\rho_0 = \Sigma / (2 H)$, and a stellar mass of $1 \, M_{\odot}$. 
To illustrate which damping terms are most important, we plot in Figure~\ref{fig:drag} the 
eccentricity damping timescale $e \, (de/dt)^{-1}$ as a function of planetary mass $M_{\rm p}$, for 
our model conditions and with $e = 0.4$ and $i = 0^{\circ}$. 
The internal density in this example was set to be $\rho_{\rm s} = 5.7 \, {\rm g} \, {\rm cm}^{-3}$,
appropriate for Earth-like bodies but slightly denser than our proto-Mars planetary embryo. 
Each curve represents damping at pericenter for different $a$.  
In Figure~\ref{fig:drag} it is seen that bodies with small mass ($M_{\rm p} < 0.1 M_{\oplus}$)
are in the drag-dominated regime, with damping timescales that increase with increasing mass as 
$M_{\rm p}^{1/3}$. 
Bodies with large mass ($M_{\rm p} > 1 M_{\oplus}$) are in the disk torque regime, with damping 
timescales that decrease as the planet's mass is increased, roughly as $M_{\rm p}^{-1}$.
The transition between these regimes, and the maximum damping timescales $\sim 10^{5} \, {\rm yr}$,
occurs for bodies of mass approximately $0.1 - 0.2 M_{\oplus}$, somewhat larger than the planetary
embryo we consider ($= 0.055 M_{\oplus}$). 

In Figure~\ref{fig:mshock_ecc} we show the eccentricity evolution of a scattered planetary embryo
with mass $M_{\rm p} = 0.055 \, M_{\oplus}$, and $R_{\rm p} = 2720 \, {\rm km}$.
The semi-major axis $a$ is set to 2.5 AU, but various starting eccentricities $e_0$ and inclinations $i_0$
for strong scattering events are explored.
These plots demonstrate that for plausible scattering scenarios the planetary embryo retains
high eccentricity for a few $\times 10^5 \, {\rm yr}$.
More relevantly, the maximum shock speeds are shown to exceed the critical value 
($\approx 6 \, {\rm km} \, {\rm s}^{-1}$) for $1 - 4 \times 10^{5} \, {\rm yr}$,
if $e_0 = 0.4$, and for a shorter time if $e_0 = 0.2$. 

Although the large eccentricities and inclinations explored here may not be common, they do represent values that could be met repeatedly during the lifetime of a disk (e.g., Chambers 2001; Hansen 2009). 


\subsection{Mass of Chondrules Produced Per Embryo\label{sec:mshock}}

We now calculate the mass of chondrules, $M_{\rm c}$, produced by a single planetary embryo.  
The rate at which chondrules are processed over time is 
$d M_{\rm c} / dt = \sigma \, V_{\rm s} \, \rho_{\rm c}$, where 
$\sigma$ is the cross section, which we set to $\pi(4300 \, {\rm km})^{2}$ for the 
planetary embryo described in \S 2. 
The density of chondrule (precursors) is defined to be $\rho_{\rm c} = f \, {\cal C} \rho_{\rm g}$, 
with $f = 3.75 \times 10^{-3}$.
The concentration factor of chondrules, ${\cal C}$, is a local value that can vary spatially
throughout the disk due to settling or turbulence.
The instantaneous velocity difference between the planetary embryo and the gas is $V_{\rm s}$,
which is also the shock speed. 
If $V_{\rm s} < 6 \, {\rm km} \, {\rm s}^{-1}$ (or greater than $8 \, {\rm km} \, {\rm s}^{-1}$)
it is assumed that chondrules are not melted (or are completely vaporized), and we set 
$d M_{\rm c} / dt = 0$ for that part of the protoplanet's orbit. 
We calculate the total mass produced by integrating $d M_{\rm c} / dt$ over individual orbits, and
over the entire damping history of the protoplanet.

Tables 1-6 list the masses of chondrules produced in bow shocks around planetary embryos with a 
variety of initial semi-major axes $a$, eccentricities $e$ and inclinations, $i$. 
Based on the need to have ${\cal C} \gtsimeq 10$ to match chondrule cooling rates (see \S 4),
we consider two situations: one in which chondrules are locally concentrated by turbulence, and
one in which chondrules have settled to the midplane.
We first consider the case where chondrules are concentrated via turbulence. 
As discussed in \S 4, chondrules more often than not find themselves in regions of enhanced chondrule
density (${\cal C} > 10$). 
As seen from the planetary embryo, which randomly  samples all volumes of the disk, the average 
density of chondrules is ${\cal C} = 1$.
Tables 1-3 list the masses of chondrules produced assuming ${\cal C} = 1$. 
We next consider the case where chondrules have settled to the midplane, by setting 
${\cal C} = 10$ if $\vert z\vert < H / 10$, and ${\cal C} = 0$ for $z$ farther from the midplane. 
Tables 4-6 list the masses of chondrules produced assuming this form of ${\cal C}$. 
It is important to note that these masses will be reduced if a large fraction of the solids
mass is locked up in large planetary bodies.
A lower solids-to-gas ratio would also result in higher chondrule cooling rates, due to lower optical depth, but probably not significantly so;  we expect them to remain within the range consistent with chondrule formation ($10 - 10^3 \, {\rm K} \, {\rm hr}^{-1}$).

As Tables 1-6 demonstrate, a single planetary embryo can produce a mass of chondrules close to
the inferred present-day mass of chondrules, $\approx 2 \times 10^{-4} \, M_{\oplus}$ (Grossman 1988). 
Higher initial eccentricities and lower initial inclinations favor greater chondrule formation
in both concentration scenarios. 
Note that when planetary bodies are scattered out of the plane of the disk, the potential chondrule-forming 
{\it duration} is greatly extended because gas drag is only effective near the orbital nodes; however, 
chondrules would only be produced at the orbital nodes, so the longer duration does not necessarily equate 
to more mass processed.   
Comparing the cases where chondrules are concentrated by turbulence vs. settled to the midplane, 
the total mass of chondrules is comparable when the planetary embryo has high inclination, whereas
if the inclination is low ($\ltsimeq 10^{\circ}$), a greater mass of chondrules is produced when
chondrules settle to the midplane.
For $a = 1.5 \, {\rm AU}$, $e_0 = 0.4$ and $i = 0^{\circ}$, $9 \times 10^{-5} \, M_{\oplus}$ 
of chondrules are produced over 64,000 years for the turbulent concentration case.  For the 
case where chondrules settle to the midplane, a factor of 10 more chondrules are produced. 

Tables 1-6 assume a planetary embryo mass $0.055 \, M_{\oplus}$. 
This may be close to the optimal mass for producing chondrules. 
Provided the body is large enough to actually produce chondrules (i.e., it is not a planetesimal
for which chondrules are accreted, but is a large planetary embryo), the cross section $\sigma$ 
scales roughly as $R_{\rm p}^{2}$.
The product of this cross section and $t_{\rm damp}$ yields the total mass of chondrules produced.
For small bodies whose damping is dominated by gas drag, the mass of chondrules that is produced 
scales linearly with $M_{\rm p}$. 
For large bodies whose damping is dominated by disk torques, the mass of chondrules that
is produced scales roughly as $M_{\rm p}^{-1/3}$. 
In fact, the mass of chondrules that can be produced per embryo is optimal for Mars-sized bodies 
with $M_{\rm p} \sim 0.1 \, M_{\oplus}$. 
Nevertheless, as long as the body is larger than about 1000 km in radius, a comparable amount
(to within factors of 3) of chondrules will be produced.

The chondrule masses produced per embryo are intriguingly close to the mass 
$2 \times 10^{-4} \, M_{\oplus}$, and in principle a single scattered planetary embryo could produce 
the entire present-day inventory of chondrules.
Nevertheless, to produce the observed mass of chondrules several planetary embryos probably
would be required.  
This is because some of the solids mass is locked up in large bodies, making fewer chondrule
precursors available for processing.
Also, the present-day asteroid belt may have been depleted since it formed, meaning there were
more chondrules originally, potentially by a factor of 100 (Weidenschilling 1977).
On the other hand, eccentric planetary embryos may be a common event.
Even for the minimum-mass solar nebula mass distribution assumed above, the annulus between
1.5 and 2.5 AU would have contained $\approx 1.5 \, M_{\oplus}$ of solids.  
If half of the mass was in small (chondrule-like) particles and half locked up in large 
planetary embryos, potentially $\approx 25$ planetary embryos could exist in that one annulus
alone.
 
 Our calculations show that, given the right orbital parameters, chondrules can form over a range 
 of heliocentric distances 1.0 - 2.5 AU. 
 Chondrules embedded in a disk with even low levels of turbulence may be spread over
 lengthscales $\Delta r = 1 \, {\rm AU}$ in $\lesssim10^{5} \, {\rm yr}$,
 assuming an alpha viscosity law $\nu = \alpha H^2 \Omega$
 (where $H$ is the gas scale height, $\Omega$ the Keplerian orbital 
 frequency and $\alpha$ a dimensionless constant $\approx 10^{-2}$;
 Hartmann et al.\ 1998).
 Chondrules produced at 1.0 - 2.5 AU will relatively quickly be spread throughout the
 present-day asteroid belt to be incorporated into chondrite parent bodies.

\section{Discussion and Summary}

Following the recent discovery that a planetary embryo, the half-formed Mars,
existed at the same time chondrules were forming (Dauphas \& Pourmond 2011), we  
have been motivated to study chondrule formation in the bow shocks of planetary 
embryos on eccentric orbits. 
We have calculated the hydrodynamics of gas flowing through the bow shock of a 
large (radius 2720 km) planetary embryo.
In contrast to bow shocks around smaller planetesimals, an embryo acquires 
a substantial primary atmosphere from the solar nebula.
A boundary layer exists between this atmosphere and the shocked gas flowing past
the planetary embryo, and this layer is unstable to KH instabilities. 
The planetary embryo is likely to have possess a magma ocean (Dauphas \& Pourmand 
2011), and the secondary atmosphere outgassed from the magma ocean would behave 
similarly dynamically. 

We have calculated the trajectories of chondrules through the planetary embryo
bow shock, subject to gravitational and gas drag forces.
We find that the relative velocity between chondrules and gas is sufficient to
melt chondrules ($> 5 \, {\rm km} \, {\rm s}^{-1}$, equivalent to shocks with 
speed $> 6 \, {\rm km} \, {\rm s}^{-1}$) only for chondrules with impact 
parameter $b < 400 \, {\rm km}$ around a 500 km radius planetesimal, or impact
parameter $b < 4300 \, {\rm km}$ around a 2720 km radius planetary embryo. 
We find that the dynamical coupling between chondrules and gas is not acheived
until the chondrules have passed about 400 km past the bow shock.  
For a small planetesimal the bow shock stands only $\approx 100$ km from the body,
and chondrules are not well coupled to the gas.  
All chondrules with $b < 460 \, {\rm km}$, including all those particles achieving
peak temperatures of chondrules and melted as chondrules, are accreted onto the 
planetesimal.
For a larger planetary embryo, the bow shock stands $\approx 1000 \, {\rm km}$ 
off of the boundary with the atmosphere, and very few chondrules are accreted 
by the protoplanet, only those with $b < 400 \, {\rm km}$ ($< 1\%$ of the chondrules
formed in the bow shock).

We have calculated the thermal histories of chondrules in a planetary embryo
bow shock.
Our calculation of thermal histories builds on the 1-D algorithm of Morris \& Desch (2010)
with the substitution of Morris et al.\ (2010a,b) for the radiation field, to allow the 
chondrules to radiate to the cold, unshocked gas.
We find that shock speeds $> 6 \, {\rm km} \, {\rm s}^{-1}$ and optical depths between
chondrules and the unshocked gas $> 0.1$ are sufficient to melt chondrules and yield
cooling rates through the crystallization temperature range $< 100 \, {\rm K} \, {\rm hr}^{-1}$,
consistent with formation of porphyritic textures. 
These optical depths, we calculate, require concentrations of chondrules ${\cal C} > 10$
above a presumed background density $3.75 \times 10^{-3} \, \rho_{\rm g}$ for a gas density
$\rho_{\rm g} = 10^{-9} \, {\rm g} \, {\rm cm}^{-3}$. 
These concentrations are achievable by chondrules settling to the midplane or
turbulent concentration. 
Moreover, a concentration factor of ${\cal C} = 10$ yields a compound chondrule frequency
in accord with observations.
To achieve the same optical depths in bow shocks around planetesimals requires much
higher concentrations of chondrules. 

We find that chondrules formed in a planetary embryo bow shock pass through the KH rolls 
at the boundary between the shocked gas and the planetary atmosphere, and will be exposed 
to the atmosphere gas.
This atmosphere is composed of nebular gas but will be dominated by volatiles outgassed
from the protoplanet's magma ocean.
We have estimated the composition of the secondary atmosphere outgassed from the protoplanet
to be roughly 10 bars of ${\rm CO}_{2}$, 3 bars of ${\rm H}_{2}{\rm O}$, and 
$2 \times 10^{-3}$ bars of Na vapor, along with other trace species.
We estimate a partial pressure of water vapor $\sim 3 \times 10^{-2}$ bar and a
partial pressure of Na vapor $\sim 2 \times 10^{-5}$ bar at the boundary region through
which chondrules traverse. 
The partial pressure of water is consistent with the abundance of fayalite in chondrules
(Fedkin et al.\ 2006) and indicators of variable oxygen fugacity in the chondrule formation
environment (Krot et al.\ 2000). 
The high partial pressures of Na may resolve the mystery of how olivines crystallized 
form chondrule melts were able to incorporate substantial Na (Alexander et al.\ 2008). 

We find that for plausibleinitial eccentricities ($e_0\gtrsim 0.2$) and inclinations
($i_0 \approx 0 $ to $ 15^{\circ}$) following a scattering event, the velocity of the planetary
embryo with respect to the gas, and the speed of the bow shock, are sufficient to melt 
chondrules and to continue to melt chondrules over the eccentricity damping timescale.
For typical chondrule-forming parameters we estimate damping timescales $\sim 10^5$ years.

Chondrules may be a byproduct of planet formation, not a necessary step toward planet formation.
Chondrule formation is likely to be delayed, perhaps by 2 Myr, until the first large
planetary embryos are formed and scattered.
Once such bodies form, several planetary embryos  can potentially melt chondrules over 
a span of several Myr, each one for $\sim 0.1 \ {\rm Myr}$ at a time.
The mass of chondrules processed by a single eccentric planetary embryo is potentially
as high as the entire mass of chondrules inferred to exist in the asteroid belt
today,  $\sim 2 \times 10^{24} \, {\rm g}$ (Grossman 1988). 
A small number of planetary embryos forming chondrules intermittently over several
Myr would be consistent with the ages of chondrules 
(Kurahashi et al. 2008; Villeneuve et al. 2009).  
A diversity of chondrule sizes, compositions and textural types could
result from the evolution of the solar nebula between chondrule-forming events 
involving protoplanets with different $a$, $e_{0}$ and $i_{0}$.

Yet another consideration is what type of object might be produced outside of the impact parameter necessary to produce chondrules.  It is likely that solids will be fully or partially melted in these regions, and experience rapid cooling rates, resulting in objects that do not fit the classical definition of chondrules.  For example, Miura et al. (2010) proposed that silicate crystals found in meteorites could be produced by supercooling of a silicate vapor in bow shocks.  Alexander et al. (2007) found that the microstructure of pyroxene in interplanetary dust particles (IDPs) indicates that it formed at temperatures $>$1258 K and cooled relatively rapidly (\app 1000 K hr$^{-1}$), suggesting some type of shock heating.  Scott \& Krot (2005) proposed that most matrix silicates likely formed as condensates  close to chondrules in transient heating events, cooling below 1300 K at \app 1000 K hr$^{-1}$.  Rubin \& Wasson (2003) suggest that ferroan olivine overgrowths on chondrules and chondrule fragments are produced by cooling rates orders of magnitude greater than those experienced by the chondrules themselves.  Formation outside the chondrule-forming impact parameter in a planetary embryo bow shock may explain objects such as the silicates found in meteorites (Scott \& Krot 2005; Miura et al. 2010), pyroxene in IDPs (Alexander et al. 2007), and overgrowths on chondrules (Wasson \& Rubin 2003; Wasson et al. 2003).  Further detailed modeling of the thermal histories of objects formed in bow shocks around planetary embryos is necessary to address this intriguing possibility. 

Finally, we speculate that there is also likely to be a critical size for a body that produces chondrules with the proper thermal histories, that are not subsequently accreted.  Based on the standoff distances found for the two bodies considered in this study (\S 3), we anticipate that this critical size is \app 1000 km in radius.  In future work, we plan a parameter study of bodies ranging in size from radii of 500 - 2720 km in order to test this hypothesis.

In this study, we have found that chondrule formation in planetary embryo bow shocks is fundamentally different from
formation in planetesimal bow shocks, but on the basis of our preliminary calculations 
this model appears to fit all the known constraints on chondrule formation. 
We intend to further develop the model, especially to include fully 2-D radiative transfer 
in determining chondrule thermal histories, to better test whether chondrules could
form in planetary bow shocks and what other types of object may form along with them.  We intend to determine the critical size of a body that is necessary for chondrule formation in bow shocks.
At the same time, we do not exclude other formation mechanisms for chondrules. 
Chondrules in CH/CB chondrites, in particular, appear to have formed from an impact
plume (Krot et al.\ 2005).
Large-scale shocks driven by gravitational instabilities also remain a viable option
for chondrule formation (Morris \& Desch 2010). 
Nevertheless, although there are some details that need to be explored further, the overall ability of the model to match chondrule constraints --- especially 
the elevated partial pressures of volatiles --- makes the planetary embryo bow shock model 
a very promising mechanism for chondrule formation. 

We gratefully acknowledge support from NASA Origins of Solar Systems 
grant NNHX10AH34G (PI SJD).
ACB's support was provided by a contract with the California Institute 
of Technology (Caltech) funded by NASA through the Sagan Fellowship Program.

\begin{table}[ht]
\begin{center}
\footnotesize
\vspace{-4 mm}
\begin{tabular}{c c c c c c}\\\hline
$a_0 \rm (AU)$ & $e_0$ & $i \rm (degrees)$ & $V_{\rm s}$ (km/s) & $M_{\rm c} (M_{\oplus})$ & Duration (yr) \\ \hline
1& 0.1& 0 & 3.0 & - & - \\
1& 0.2& 0 & 6.1  & 8.3(-7) & 1 000 \\
1& 0.3& 0 & 9.4  & 7.4(-5) & 21 000 \\
1& 0.4& 0 & 13 & 1.1(-4) &  29 000 \\\hline

1& 0.1& 7 & 4.8 & - & -\\
1& 0.2& 7 & 7.2 & 2.5(-5) & 12 000\\
1& 0.3& 7 & 10.2 & 6.5(-5) & 24 000\\
1& 0.4& 7 & 13.7 & 7.4(-5) & 22 000\\\hline

1& 0.1& 15 & 8.4 & 4.6(-5) & 43 000\\
1& 0.2& 15 & 10.1 &  3.7(-5) & 31 000\\
1& 0.3& 15 & 12.5 & 3.6(-5) & 25 000\\
1& 0.4& 15 & 15.6 & 4.3(-5) & 22 000\\\hline\hline
\end{tabular}
\end{center}
\caption{\footnotesize 
Masses of chondrules produced by planetary embryo bow shock over the orbital evolution of the
body, assuming an initial semi-major axis of $a = 1.0 \, {\rm AU}$ and different initial values 
eccentricity $e$ and inclination $i$.  
We assume chondrule formation occurs only when the instantaneous velocity difference $V_{\rm s}$ 
between the embryo and the gas is in the range $6 - 8 \, {\rm km} \, {\rm s}^{-1}$. 
Max $V_{\rm s}$ is the maximum shock velocity during the orbital evolution. 
The duration signifies the time period beginning with the first episode of chondrule formation 
and ending with the last episode of chondrule formation before damping of the embryo's $e$ and $i$. 
A concentration factor ${\cal C} = 1$ (implying chondrules-to-gas mass ratio 0.00375) was assumed.
Only material that is within $H/C$, for disk gas scale height $H$ is included in $M_s$. 
The notation $a(-b)$ means $a\times10^{-b}$.}
\vspace{-3 mm}
\end{table}

\begin{table}[ht]
\begin{center}
\footnotesize
\vspace{-4 mm}
\begin{tabular}{c c c c c c}\\\hline
$a_0 \rm (AU)$ & $e_0$ & $i \rm (degrees)$ & $V_{\rm s}$ (km/s) & $M_{\rm c} (M_{\oplus})$ & Duration (yr) \\\hline

1.5& 0.1& 0 & 2.5 & - & -\\
1.5& 0.2& 0 & 5.0 & - & -\\
1.5& 0.3& 0 & 7.7 & 3.7(-5) & 32 000\\
1.5& 0.4& 0 & 11 & 8.8(-5) & 64 000\\\hline

1.5& 0.1& 7 & 3.9 & - & -\\
1.5& 0.2& 7 & 5.9 & - & -\\
1.5& 0.3& 7 &  8.3 & 5.1(-5) & 55 000\\
1.5& 0.4& 7 & 11 & 6.5(-5)&  56 000  \\\hline

1.5& 0.1& 15 & 6.9 & 1.3(-5) & 37 000\\
1.5& 0.2& 15 & 8.2 & 2.7(-5) &  68 000\\
1.5& 0.3& 15  & 10 & 2.9(-5) & 63 000\\
1.5& 0.4& 15 & 13 & 3.6(-5) & 58 000\\\hline\hline
\end{tabular}
\end{center}
\caption{\footnotesize Same as Table 1, but for $a = 1.5 \, {\rm AU}$.}
\vspace{-3 mm}
\end{table}

\begin{table}[ht]
\begin{center}
\footnotesize
\vspace{-4 mm}
\begin{tabular}{c c c c c c}\\ \hline
$a_0 \rm (AU)$ & $e_0$ & $i \rm (degrees)$ & $V_{\rm s}$ (km/s) & $M_{\rm c} (M_{\oplus})$ & Duration (yr) \\ \hline

2.5& 0.1& 0 & 1.9 & - & -\\
2.5& 0.2& 0 & 3.9 & - & -\\
2.5& 0.3& 0 & 5.97 & - & -\\
2.5& 0.4& 0 & 8.3 & 5.5(-5) & 140 000\\\hline

2.5& 0.1& 7 & 3.0 & - & -\\
2.5& 0.2& 7 & 4.5 & - & -\\
2.5& 0.3& 7 &  6.4 & 5.5(-6) & 35 000\\
2.5& 0.4& 7 & 8.7 & 5.5(-5)&  180 000  \\\hline

2.5& 0.1& 15 & 5.3 & - & -\\
2.5& 0.2& 15 & 6.4 & 3.1(-6) &  32 000\\
2.5& 0.3& 15  & 7.9 & 2.2(-5) & 180 000\\
2.5& 0.4& 15 & 9.9 & 2.8(-5) & 190 000\\\hline\hline
\end{tabular}
\end{center}
\caption{\footnotesize Same as Table 1, but for $a = 2.5 \, {\rm AU}$.}
\vspace{-3 mm}
\end{table}

\begin{table}[ht]
\begin{center}
\footnotesize
\vspace{-4 mm}
\begin{tabular}{c c c c c c}\\\hline
$a_0 \rm (AU)$ & $e_0$ & $i \rm (degrees)$ & $V_{\rm s}$ (km/s) & $M_{\rm c} (M_{\oplus})$ & Duration (yr) \\\hline
1& 0.1& 0 & 3.0 & - & - \\
1& 0.2& 0 & 6.1  & 8.3(-6) & 1 000 \\
1& 0.3& 0 & 9.4  & 7.4(-4) & 21 000 \\
1& 0.4& 0 & 13 & 1.1(-3) &  29 000 \\\hline

1& 0.1& 7 & 4.8 & - & -\\
1& 0.2& 7 & 7.2 & 3.0(-5) & 12 000\\
1& 0.3& 7 & 10.2 & 6.5(-5) & 17 000\\
1& 0.4& 7 & 13.7 & 7.9(-5) & 12 000\\\hline

1& 0.1& 15 & 8.4 & 4.6(-5) & 40 000\\
1& 0.2& 15 & 10.1 &  3.6(-5) & 28 000\\
1& 0.3& 15 & 12.5 & 3.6(-5) & 22 000\\
1& 0.4& 15 & 15.6 & 4.3(-5) & 18 000\\\hline\hline
\end{tabular}
\end{center}
\caption{\footnotesize Same as Table 1, but assuming a concentration of chondrules 
${\cal C} = 10$ within a height $\vert z\vert < H/10$ of the midplane. }
\vspace{-3 mm}
\end{table}

\begin{table}[ht]
\begin{center}
\footnotesize
\vspace{-4 mm}
\begin{tabular}{c c c c c c}\\\hline
$a_0 \rm (AU)$ & $e_0$ & $i \rm (degrees)$ & $V_{\rm s}$ (km/s) & $M_{\rm c} (M_{\oplus})$ & Duration (yr) \\\hline

1.5& 0.1& 0 & 2.5 & - & -\\
1.5& 0.2& 0 & 5.0 & - & -\\
1.5& 0.3& 0 & 7.7 & 3.7(-4) & 32 000\\
1.5& 0.4& 0 & 11 & 8.8(-4) & 64 000\\\hline

1.5& 0.1& 7 & 3.9 & - & -\\
1.5& 0.2& 7 & 5.9 & - & -\\
1.5& 0.3& 7 &  8.3 & 6.9(-5) & 47 000\\
1.5& 0.4& 7 & 11 & 7.9(-5)&  33 000  \\\hline

1.5& 0.1& 15 & 6.9 & 1.5(-5) & 37 000\\
1.5& 0.2& 15 & 8.2 & 2.8(-5) &  62 000\\
1.5& 0.3& 15  & 10 & 2.9(-5) & 56 000\\
1.5& 0.4& 15 & 13 & 3.6(-5) & 49 000\\\hline\hline
\end{tabular}
\end{center}
\caption{\footnotesize Same as Table 4, but with $a = 1.5 \, {\rm AU}$. }
\vspace{-3 mm}
\end{table}

\begin{table}[ht]
\begin{center}
\footnotesize
\vspace{-4 mm}
\begin{tabular}{c c c c c c}\\\hline
$a_0 \rm (AU)$ & $e_0$ & $i \rm (degrees)$ & $V_{\rm s}$ (km/s) & $M_{\rm c} (M_{\oplus})$ & Duration (yr) \\\hline

2.5& 0.1& 0 & 1.9 & - & -\\
2.5& 0.2& 0 & 3.9 & - & -\\
2.5& 0.3& 0 & 5.97 & - & -\\
2.5& 0.4& 0 & 8.3 & 5.5(-4) & 140 000\\\hline

2.5& 0.1& 7 & 3.0 & - & -\\
2.5& 0.2& 7 & 4.5 & - & -\\
2.5& 0.3& 7 &  6.4 & 8.8(-6) & 35 000\\
2.5& 0.4& 7 & 8.7 & 7.4(-5)&  130 000  \\\hline

2.5& 0.1& 15 & 5.3 & - & -\\
2.5& 0.2& 15 & 6.4 & 3.5(-6) &  32 000\\
2.5& 0.3& 15  & 7.9 & 2.4(-5) & 180 000\\
2.5& 0.4& 15 & 9.9 & 2.9(-5) & 170 000\\\hline\hline
\end{tabular}
\end{center}
\caption{\footnotesize Same as Table 4, but with $a = 2.5 \, {\rm AU}$. }
\vspace{-3 mm}
\end{table}

\begin{figure}
\includegraphics[width=15cm]{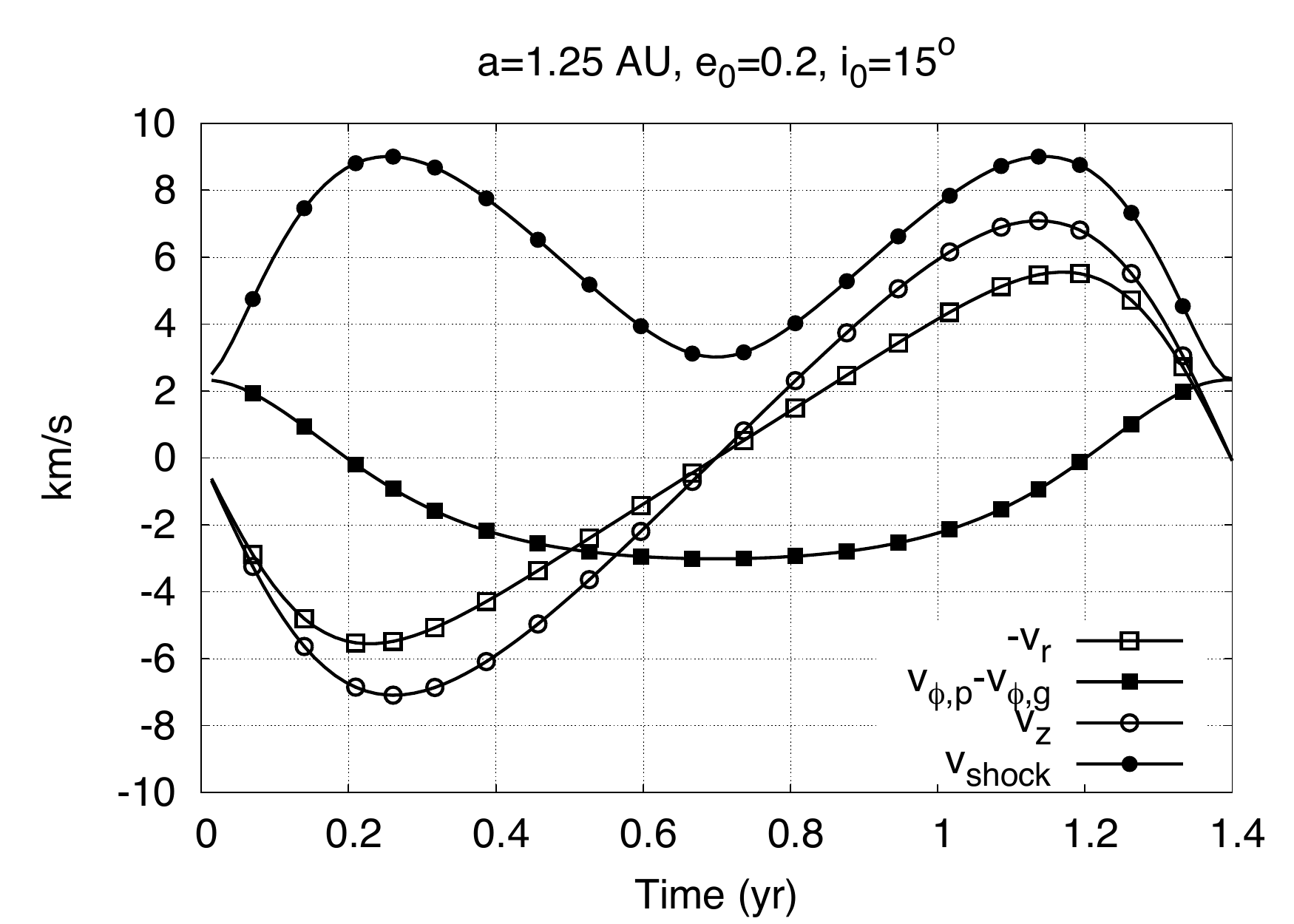}
\caption{Velocity difference $V_{\rm s}$ between nebular gas (on a circular Keplerian orbit)
and a planetary embryo on an eccentric orbit with semi-major axis $a = 1.25 \, {\rm AU}$, eccentricity 
$e = 0.2$ and inclination $i = 15^{\circ}$.
The radial, azimuthal and vertical components, and the total velocity difference $V_{\rm s}$
(equal to the speed of the bow shock) are shown as functions of time over the orbit, where $t = 0$
signifies perihelion passage.  
In this example the planet passes through the ascending node at $t \approx 1.1$ years.
$V_{\rm s}$ is supersonic and bow shocks are driven throughout the orbit. 
\label{fig:1}}
\end{figure}

\begin{figure}
\includegraphics[width=15cm]{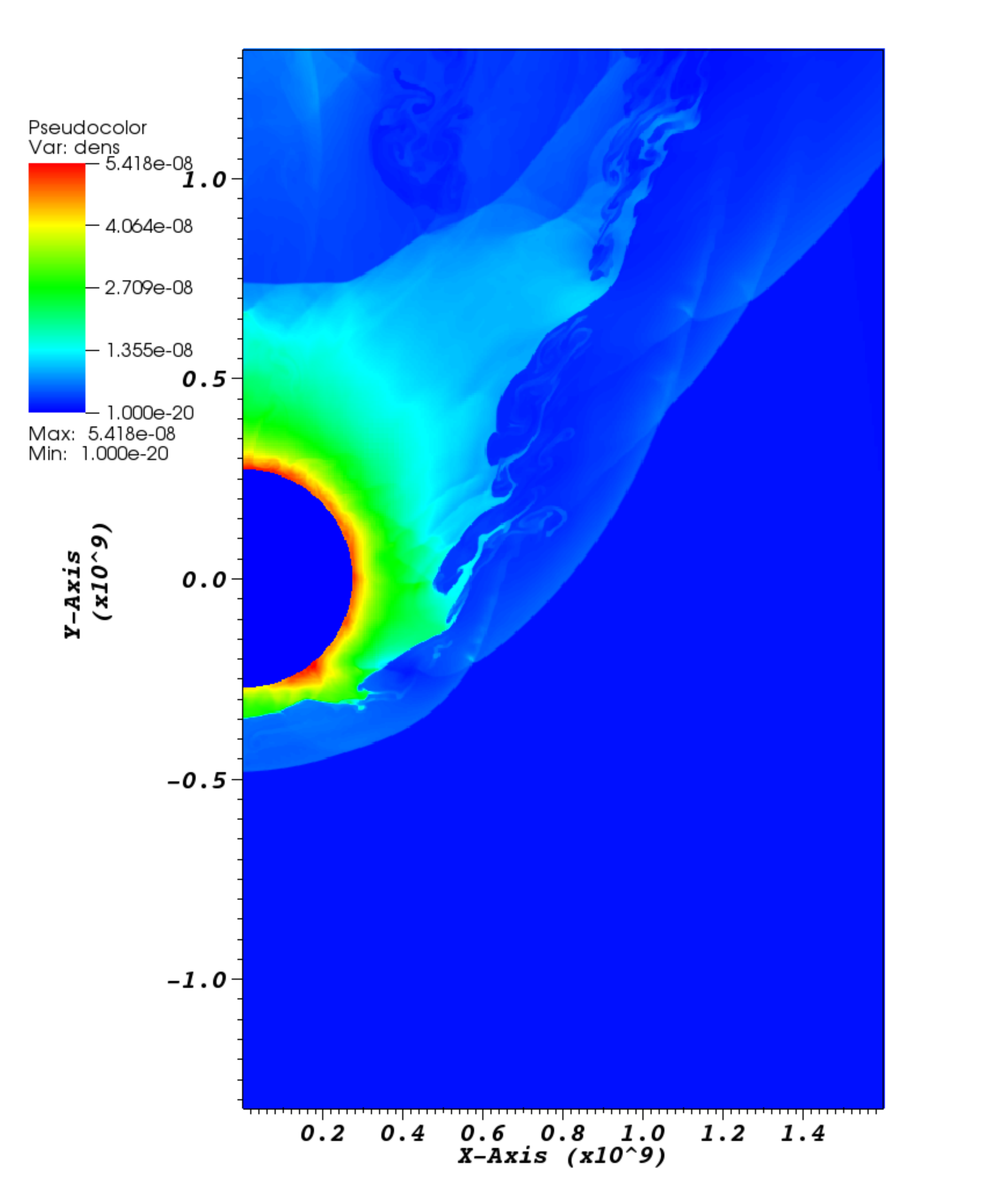}
\caption{Animation showing the evolution of a bow shock around a planetary embryo with radius 2720 km (i.e. half a Mars mass), on an inclined and eccentric orbit.
\label{movie}}
\end{figure}

\begin{figure}
\includegraphics[width=8cm]{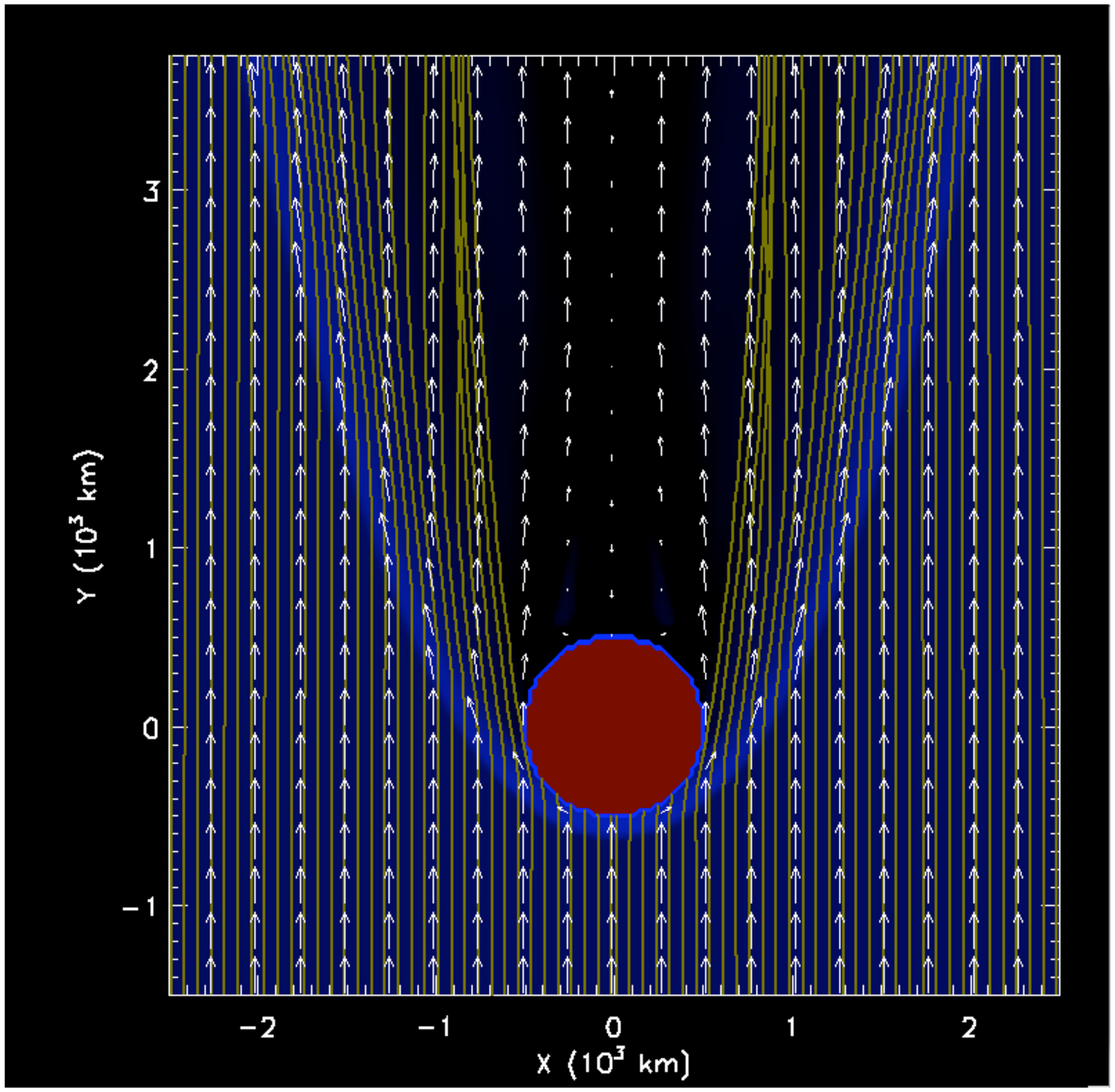}\includegraphics[width=8cm]{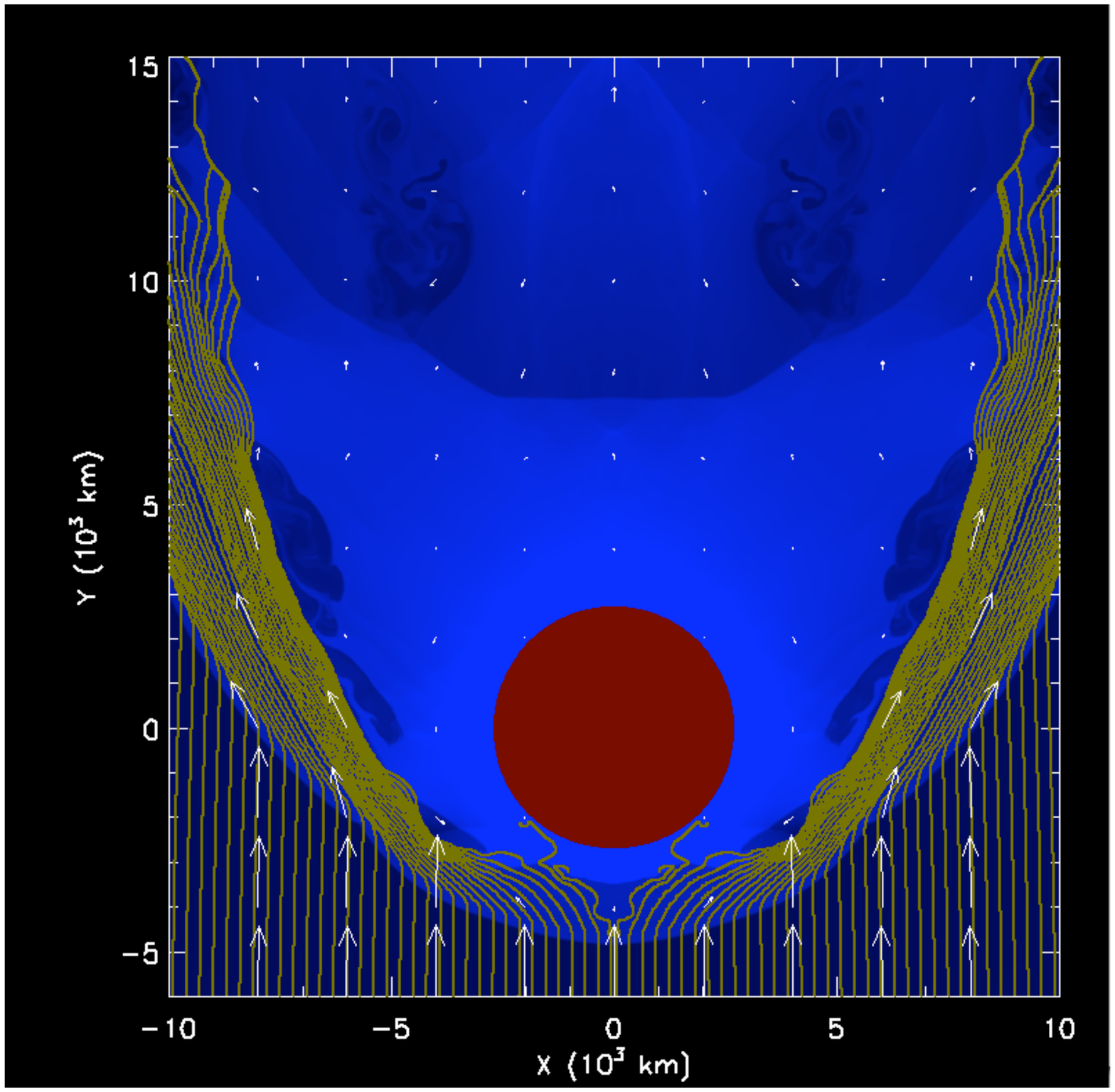}
\caption{Left: Gas density (blue contours) and velocities (white arrows),
and trajectories of chondrule precursors / chondrules (gold streamlines),
in the vicinity of the bow shock surrounding a 500-km radius planetesimal.
Velocities are measured in the frame of the planetesimal, moving at 
$8 \, {\rm km} \, {\rm s}^{-1}$ with respect to the local gas.  Chondrules 
with impact parameters $\lesssim$ 460 km are accreted.
Right: The same plot repeated for a planetary embryo with radius
2720 km (i.e., half a Mars mass).  Note the change in scale.  In this 
case, particles with impact parameter $\gtrsim$ 400 km are 
{\it not} accreted. \label{fig:streamlines}}
\end{figure}

\begin{figure}
\includegraphics[width=6.5cm,angle=-90]{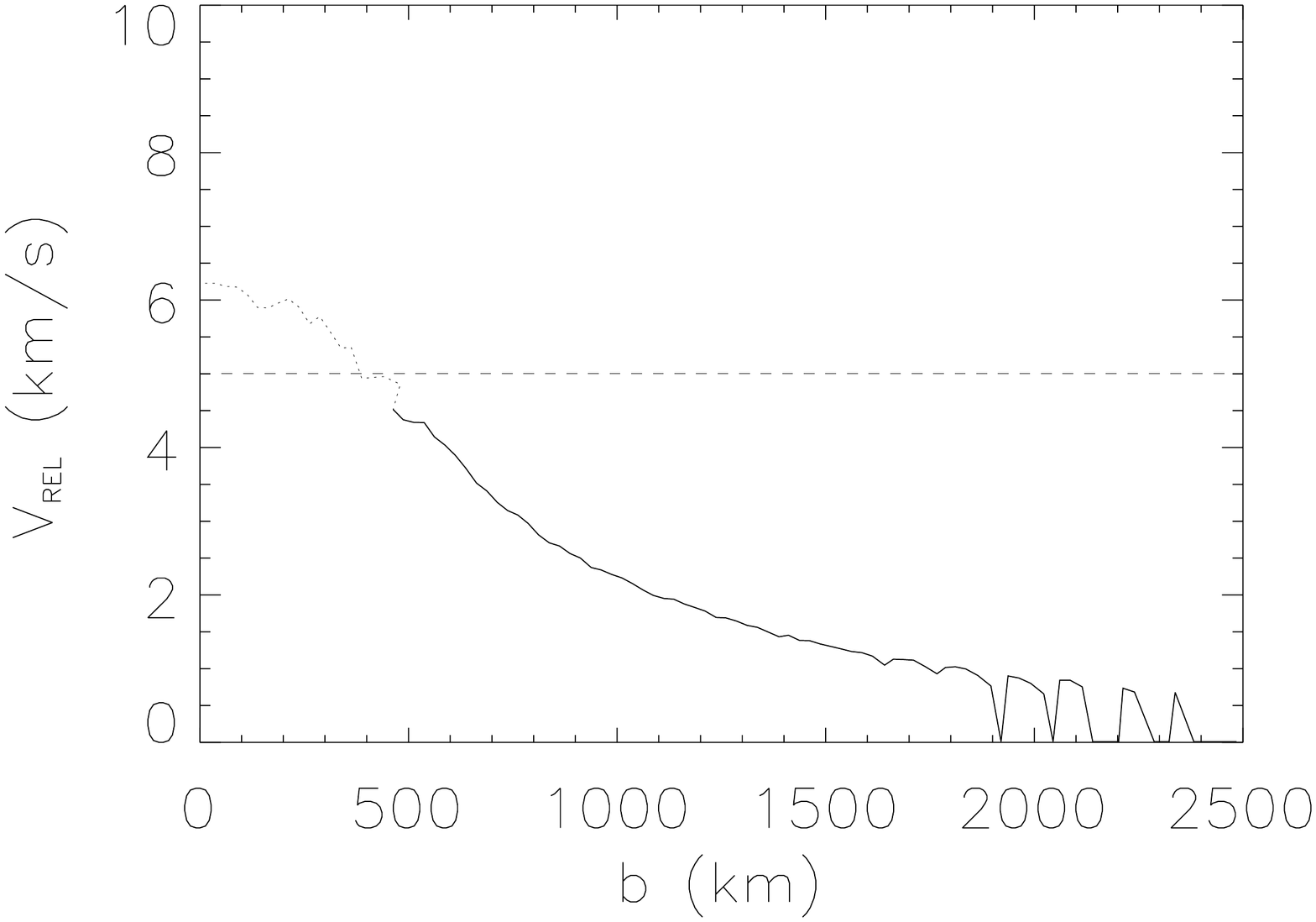}\includegraphics[width=6.5cm,angle=-90]{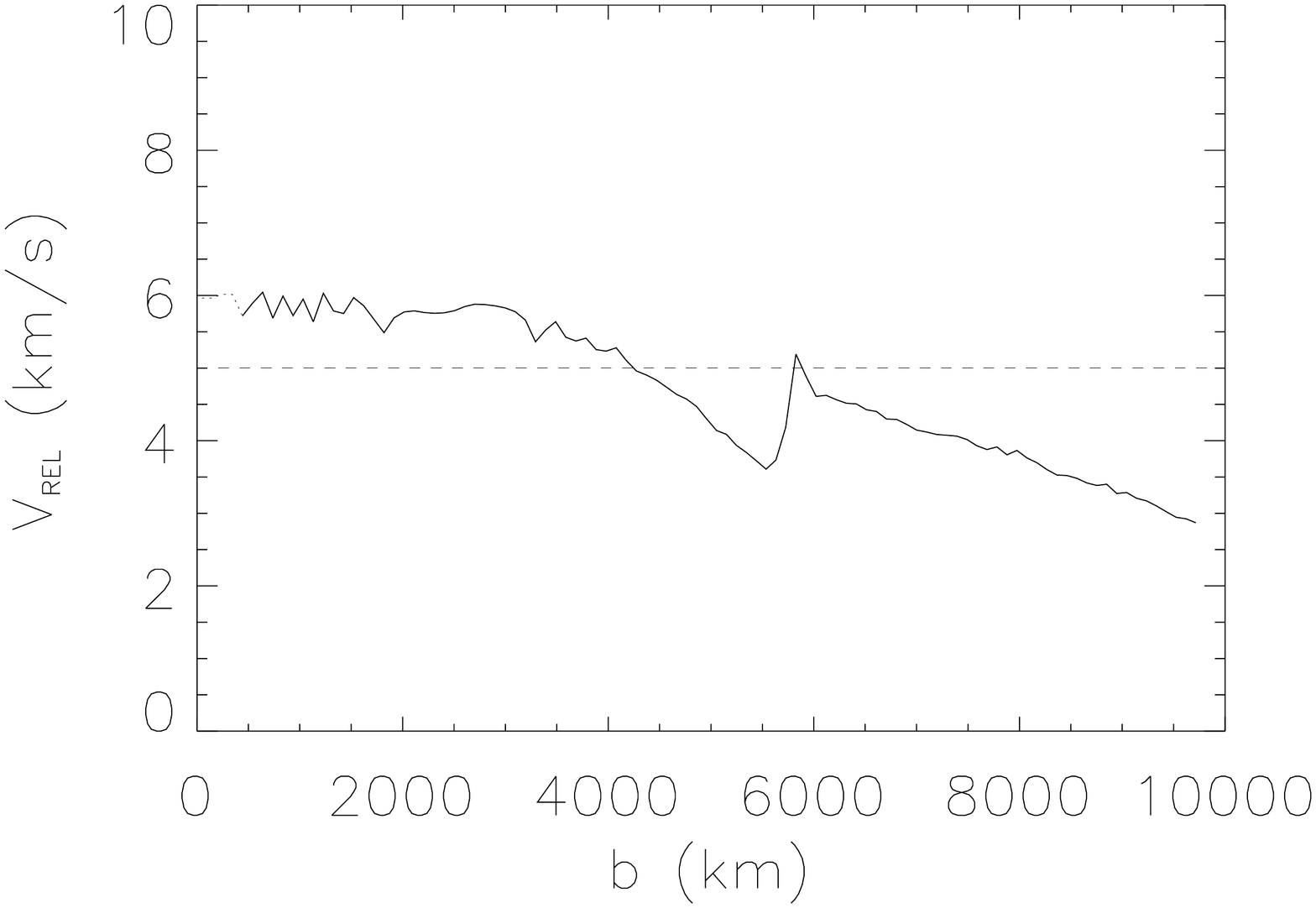}
\caption{Left: Maximum relative velocity between chondrules and gas,
achieved just after passing through the bow shock, as a function of
the impact parameter $b$, for the $R_{\rm p} = 500 \, {\rm km}$ case.  
Values of $b$ that lead to the chondrules being accreted are plotted
with a dotted line instead of a solid line.  The minimum relative 
velocity needed for chondrules to melt ($= 5 \, {\rm km} \, {\rm s}^{-1}$)
is shown.
Right: The same but for the $R_{\rm p} = 2720 \, {\rm km}$ case.
Only particles with $b \lesssim 400 \, {\rm km}$ might be accreted.} 
\label{fig:vel}
\end{figure}

\begin{figure}
\includegraphics[width=8.75cm]{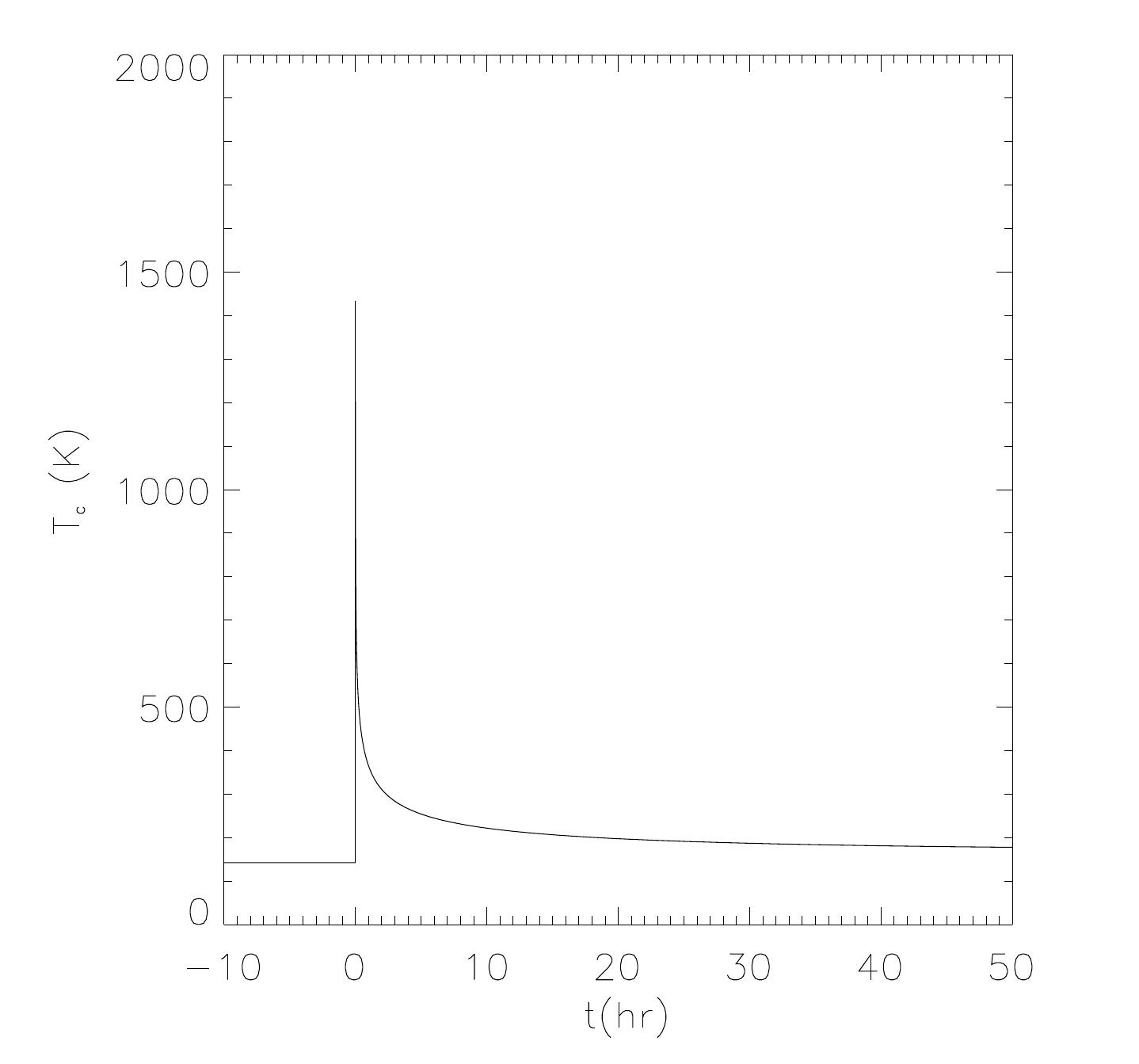}\includegraphics[width=8.75cm]{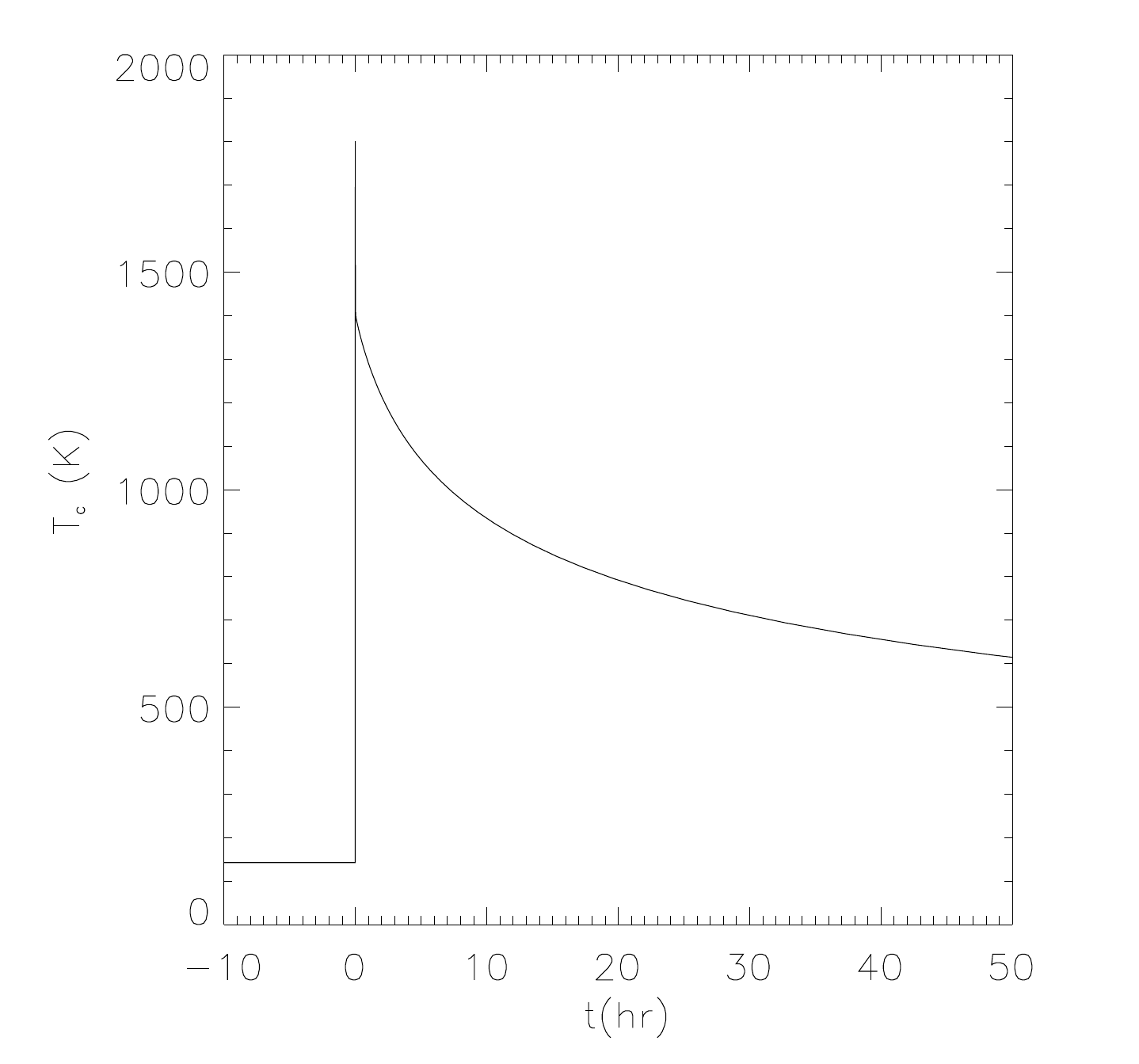}
\includegraphics[width=8.75cm]{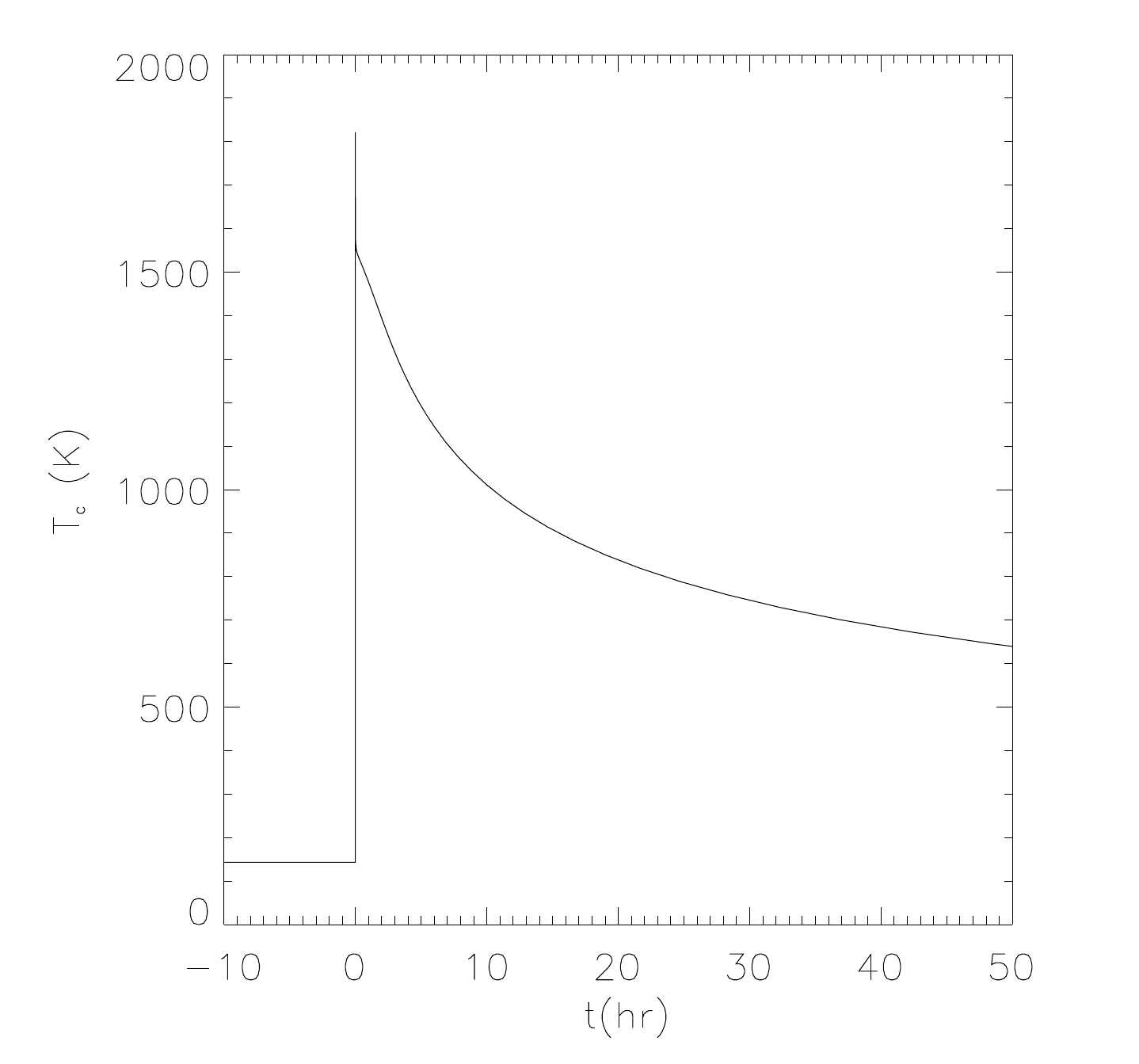}\includegraphics[width=8.75cm]{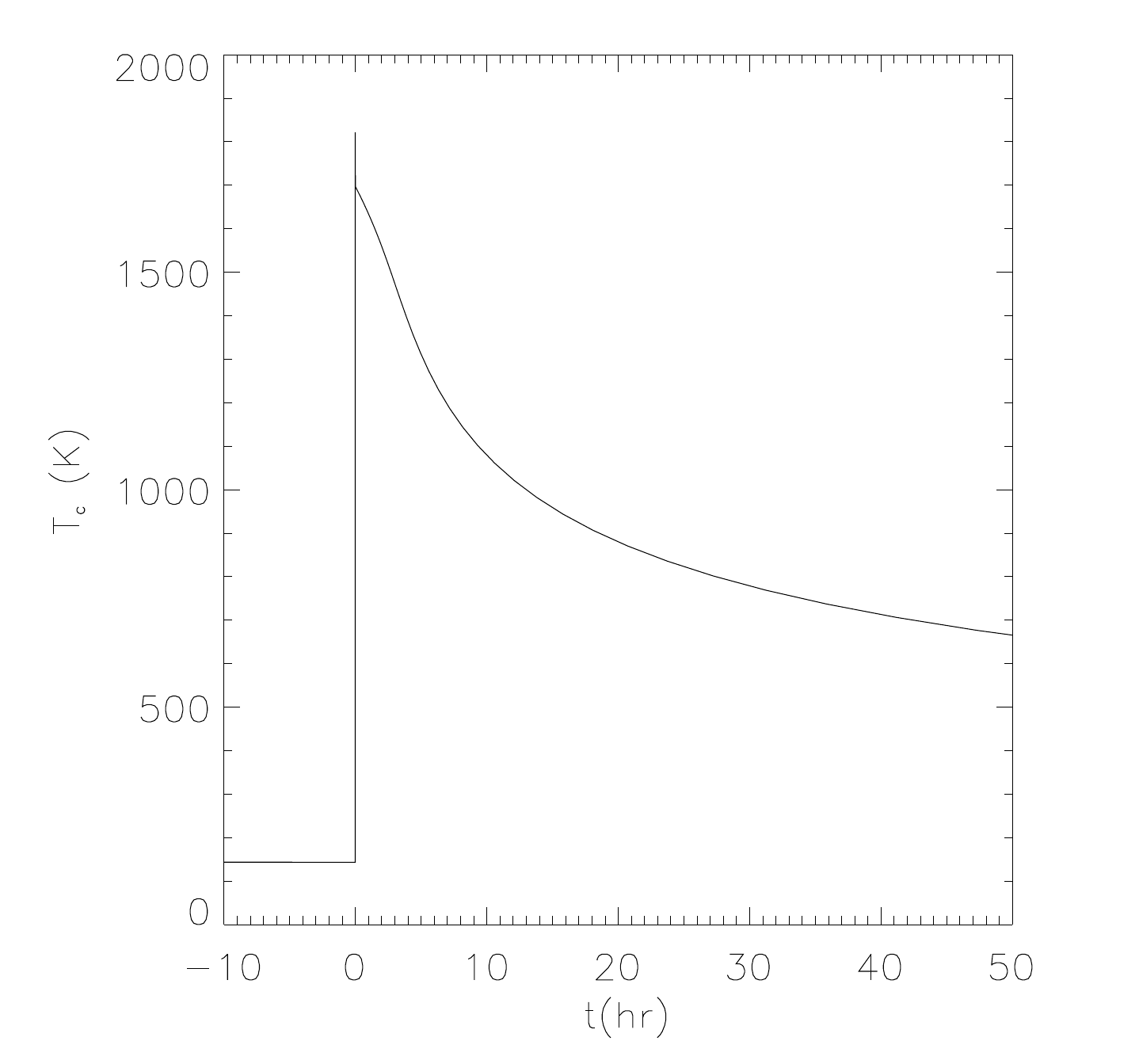}
\caption{Thermal histories of chondrules formed in bow shocks around planetary embryos, assuming shock speeds
of $5 \, {\rm km} \, {\rm s}^{-1}$ (top left), $6 \, {\rm km} \, {\rm s}^{-1}$ (top right), $7 \, {\rm km} \, {\rm s}^{-1}$ (bottom left), and 
$8 \, {\rm km} \, {\rm s}^{-1}$ (bottom right).   Chondrule precursors pass through the shock front at $t = 0$.
See text for details. \label{fig:cool}}
\end{figure}

\begin{figure}
\includegraphics[width=15cm]{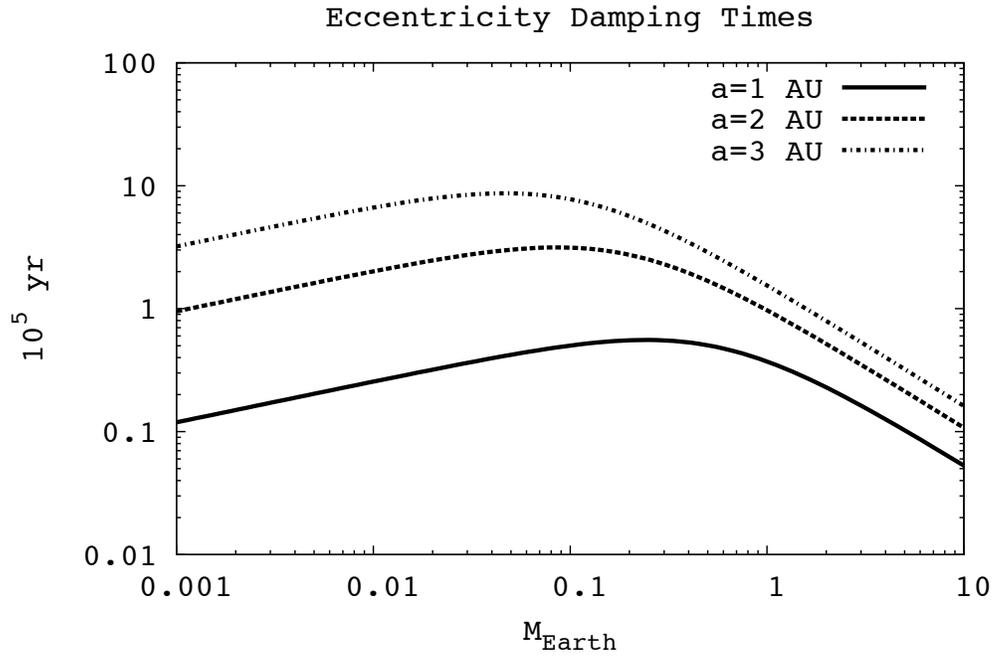}
\caption{Eccentricity damping times as a function of planetary embryo mass.  Damping for small
bodies ($M < 0.1 \, M_{\oplus}$) is dominated by gas drag, and damping of large bodies 
($M > 1 \, M_{\oplus}$) is dominated by disk torques.  For heliocentric distances 1 - 3 AU,
Mars-sized bodies are in the transition regime that is least damped. }
\label{fig:drag}
\end{figure}

\begin{figure}
\includegraphics[width=8cm]{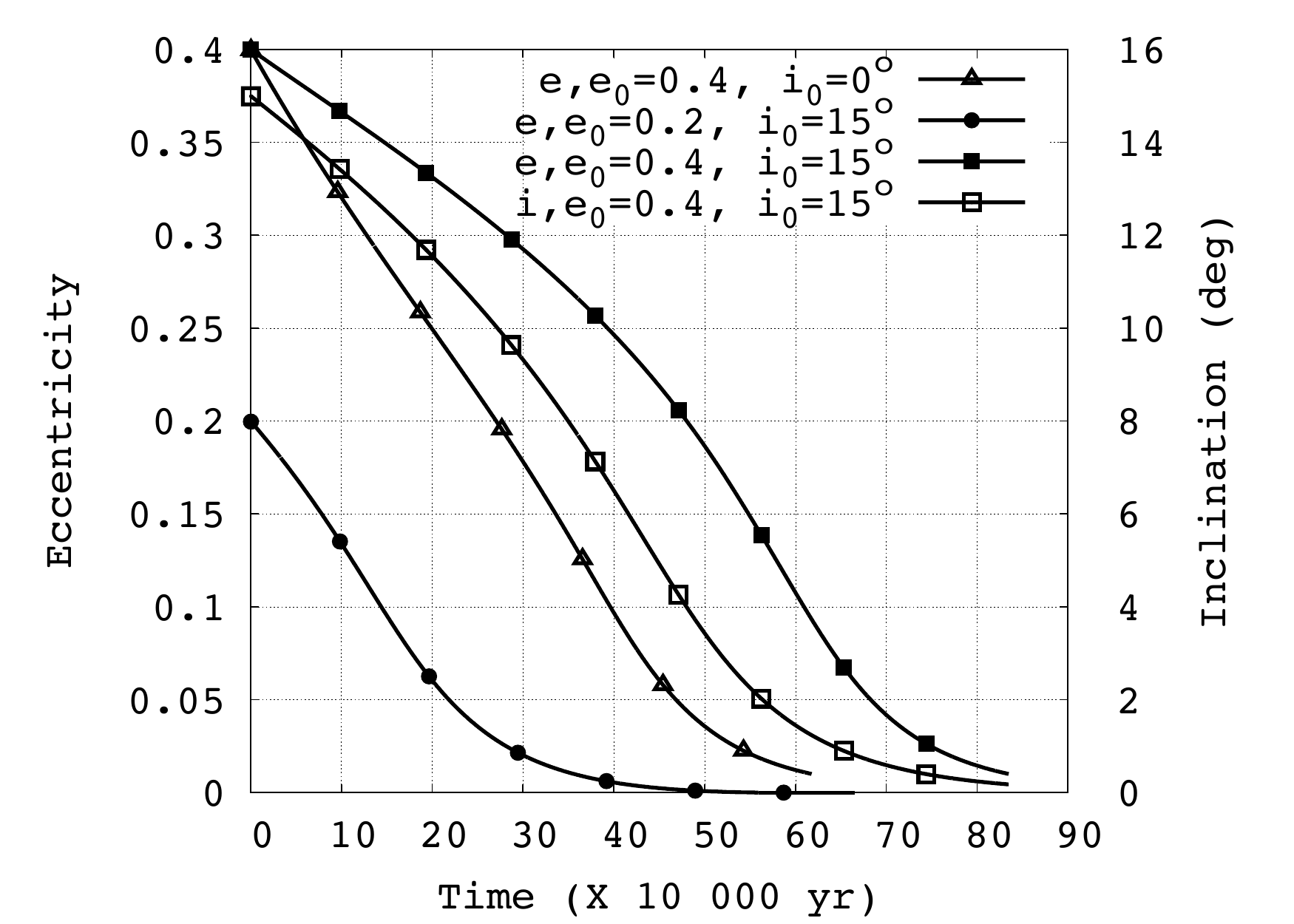}\includegraphics[width=8cm]{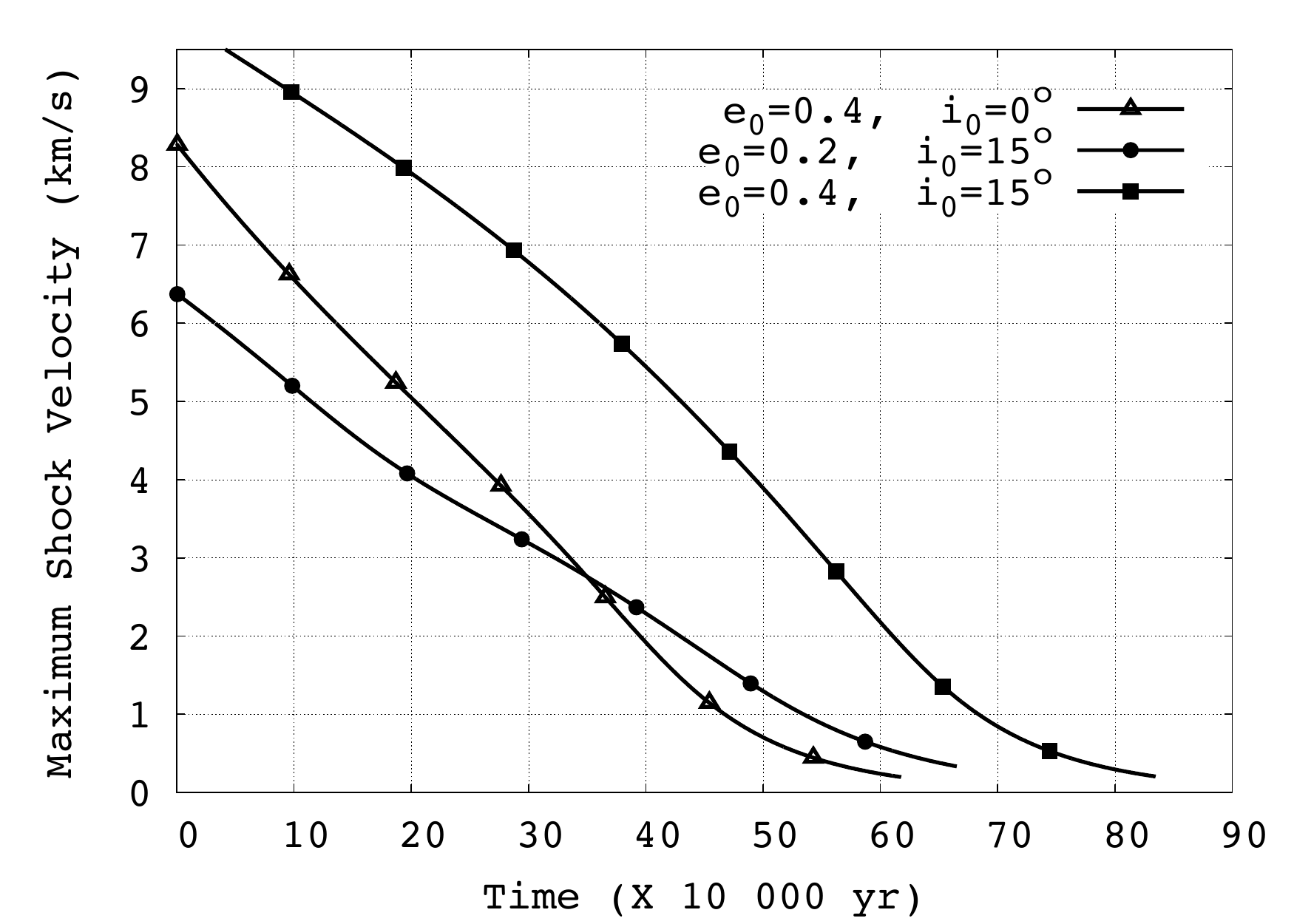}
\caption{Left: Eccentricity evolution for three scattered-planet scenarios.  
Each embryo has mass $0.055 \, M_{\oplus}$, radius 2720 km, and semi-major axis 2.5 AU.
Curves correspond to starting conditions $e_0 = 0.4$, $i_0 = 0^{\circ}$ (solid triangles),
$e_0 = 0.2$, $i_0 = 15^{\circ}$ (solid circles)
$e_0 = 0.4$, $i_0 = 15^{\circ}$ (solid and open squares).
Right: Maximum shock velocities during an orbit, as a function of time, for the same three cases. 
\label{fig:mshock_ecc}}
\end{figure}

\label{lastpage}

\end{document}